\documentclass[%
reprint,
amsmath,amssymb,
aps,
prd,
]{revtex4-1}

\def\nat{\rm{Nature~}}           
\def\icarus{\rm{Icarus~}}           
\def\mnras{\rm{Mon. Not. R. Astron. Soc.~}}             
\def\jpe{\rm{J. Phys. Earth~}}      
\def\aap{\rm{Astron. Astrophys.~}}      
\def\apj{\rm{Astrophys.~J.~}}      
\def\apjl{\rm{Astrophys.~J.~Lett.~}}      
\def\plb{\rm{Phys.~Lett.~B~}}        
\def\physrep{\rm{Phys.~Rep.~}}        
\def\prc{\rm{Phys.~Rev.~C }}        
\def\prd{\rm{Phys.~Rev.~D }}        
\def\prl{\rm{Phys.~Rev.~Lett. }}    
\usepackage{graphicx}
\usepackage{dcolumn}
\usepackage{amsmath}
\usepackage{amssymb}
\usepackage{float}
\usepackage{multirow}

\usepackage{epstopdf}
\graphicspath{ {Figures/} }


\allowdisplaybreaks

\begin{document}
	
	
	\title{Two-layer compact stars with crystalline quark matter: Screening effect on the tidal 
		deformability }
	
	
	\author{S. Y. Lau}
	\altaffiliation[Email address: ]{sylau@phy.cuhk.edu.hk}
	\author{P. T. Leung}
	\altaffiliation[Email address: ]{ptleung@phy.cuhk.edu.hk}
	\author{L.-M. Lin}%
	\altaffiliation[Email address: ]{lmlin@phy.cuhk.edu.hk}
	\affiliation{%
		Department of Physics, The Chinese University of Hong Kong, Hong Kong, China
	}%
	
	
	\date{\today}
	
	\begin{abstract}
		It is well known that the tidal deformability of a compact star carries important information about the interior equation-of-state (EOS) of the star. The first gravitational-wave event GW170817 from a binary compact star merger observed by the LIGO/VIRGO detectors have already put limits on the tidal deformability and provided constraints on the ultra-high nuclear density EOS. In view of this ground breaking discovery, we revisit and extend our previous work [Phys. Rev. D {\bf 95}, 101302(R) (2017)] which found that taking the effect of elasticity into account in the calculation of the tidal deformability of compact star models composed of crystalline color-superconducting (CCS) quark matter can break the universal I-Love relation discovered for fluid compact stars. In this paper, we present our formulation in detail and provide more analysis to complement our previous findings. We focus and extend the study of the screening effect on the tidal deformability, which we found previously for hybrid star models, to various theoretical two-layer compact star models. Besides solid quark stars and hybrid stars, we also consider (1) solid quark stars dressed in a thin nuclear matter crust and (2) quark stars with a fluid quark-matter core in the color-flavor-locked phase surrounded by a solid CCS quark matter envelope. We show that the screening effect of these two-layer models in general depends on the thickness of the envelope and the ratio between the density gap and the core density at the core-envelope interface. However, for models with a 
		fluid envelope and a vanishing small density gap, 
		the screening effect remains strong even as the thickness of the envelops tends to 
		zero if the quark matter core has a fairly uniform density.		
		The relevance of our study to GW170817 is also discussed. We find that quark star models which are ruled out by the observation limits on the tidal deformability can be revived if the entire quark star is in a CCS phase instead of a fluid phase, thus adding complication on putting constraints on the quark star EOSs. In contrast, the screening effect causes the tidal deformability of a hybrid star with a CCS quark matter core agrees with that of a corresponding stellar model with a fluid core to within less than 1\% if the core size is less than about 70\% of the stellar radius. The implication is that if a hybrid star EOS model is ruled out by the observation limits on the tidal deformability, the conclusion will hold no matter whether the quark matter is in a fluid or solid state, assuming that a large solid core comparable to the stellar radius is not favored in nature.
		Our study advocates that the tidal deformability not only provides us information on the EOS, but may also give insights into the multi-layer structure and elastic properties of compact star models composed of CCS quark matter.
	\end{abstract}
	
	\pacs{Valid PACS appear here}
	\maketitle
	
	
	%
	%
	%
	%
	%
	
	\section{\label{sec1:level1} Introduction}
	
	Compact stars have long been perceived as natural laboratories of matter in extremely high density and low temperature, which cannot be attained on Earth. The uncertainties of the equations of state (EOSs) of matter in such an environment can be constrained through observing the signals from compact stars. Aside from traditional observations of electromagnetic signals, gravitational wave signals from compact stars first come into play last year. On 17 August 2017, the Advanced LIGO and Virgo network made the first successful detection of the gravitational wave signal from a binary compact star system, GW170817 \cite{PhysRevLett.119.161101}. The first analysis of the signal already places upper bounds on the tidal deformability, which is a parameter quantifying the ratio of the induced quadrupole moment of a star to an external tidal field. This parameter,  denoted by $\lambda$, is shown to be encoded in the emitted gravitational wave signals as a small correction in the phase of the waveform during the early stage of binary compact star inspirals. The measurability of $\lambda$ with gravitational-wave observations has been studied \cite{2008PhRvD..77b1502F,2012PhRvD..85l3007D,PhysRevLett.111.071101,2013PhRvD..88d4042R,PhysRevLett.112.101101,PhysRevD.89.021303,2015PhRvD..91d3002L,2016PhRvD..93f4082H,2017PhRvD..95j4036L,2017PhRvD..96l1501D,2017PhRvD..96h4060K}. Since this parameter is sensitive to the EOS \cite{2009PhRvD..80h4035D,2010PhRvD..82b4016P}, the gravitational wave signals effectively carries information of the EOS independent of the electromagnetic signals. Indeed, based on the GW170817 signal, work has been done to put constraints on the compact star EOSs (\cite{2018ApJ...852L..29R,PhysRevLett.120.172703,PhysRevD.97.083015,PhysRevD.97.084038,2018ApJ...857L..23R,2018ApJ...852L..25R,2017ApJ...850L..34B,PhysRevLett.120.172702,2018RAA....18...24L,Abbott:2018exr}). With this ground breaking first observation, the prospect of using the tidal deformability to probe the properties of compact stars in the future is very promising.
	
	One of the open questions that may be explored with gravitational wave measurements is the properties of quark matter in a cold dense environment.
	It is generally believed that deconfined quarks may exist inside the core of compact stars, which corresponds to the high density and low temperature region of the QCD phase diagram \cite{1965Ap......1..251I, 1970PThPh..44..291I,1973PhLB...47..365F,1975PhRvL..34.1353C, 1976PhLB...62..241B}. At asymptotically high density, QCD predicts that the up, down and strange quarks in the deconfined quark matter would pair up equally to form standard Cooper pairs based on the BCS mechanism and become color superconducting. This phase of quark matter is called the color-flavor-locked (CFL) phase \cite{1999NuPhB.537..443A}.
	
	On the other hand, the phase of quark matter at a relatively lower density within a compact star is uncertain when the strange quark mass ($\sim 150~\text{MeV}$) is comparable to the quark chemical potential ($\sim 400~\text{MeV}$) since perturbative QCD is no longer adequate in this regime. It is proposed that in such conditions quark matter may be in the crystalline color superconducting (CCS) phase \cite{2001afpp.book.2061R,2001PhRvD..63g4016A,2005PhLB..627...89C,2006PhRvD..73k4012M,2006PhRvD..74i4019R,2006PhLB..642..350C}, which is a rigid state of matter expected to have an extremely high shear modulus about 20-1000 times that of nuclear matter in a neutron star crust \cite{2007PhRvD..76g4026M}. The exact transition point between the CFL phase and the CCS phase is highly uncertain and the possibility of such a transition existing within a compact star cannot be ruled out \cite{2014PhRvD..89j3014M,2015ApJ...815...81M}. The possibility of sequential QCD phase transitions within compact stars has also been studied \cite{2017PhRvL.119p1104A}.
	
	Moreover, there might be a transition point between the quark matter phase to nuclear matter lying within the density range of a compact star. As a result, several phases of matter with distinct properties could exist within compact stars. For instance, `hybrid stars' containing a quark matter core and a nuclear matter envelope has long been hypothesized to be some of the observed compact stars \cite{1975PhRvL..34.1353C,1976PhLB...62..241B,1976Natur.264..235C,2000PhR...328..237H}.
	
	In another scenario where strange matter is the absolute ground state for strong interactions \cite{1984PhRvD..30..272W}, most of the hadronic matter is turned into deconfined quark matter within a compact star. A thin layer of nuclear matter might exist on the top of the quark matter since the compact star attracts normal nuclear matter from the surroundings. The strange quark matter is not in direct contact with the nuclear matter crust due to Coulomb repulsion. As a result, such a model is composed of quark matter dressed in a thin layer of nuclear matter. The two phases of matter are separated by a thin layer of electrons \cite{1986ApJ...310..261A,2015ApJ...815...81M}.
	
	While different theoretical possibilities have been proposed, could we tell from observations in what phase(s) deconfined quark matter (if exists) can occur in compact stars? This is certainly 
	a non-trivial question since even the EOS of traditional neutron stars is 
	still an open question. Following the successful measurement of the gravitational wave signal GW170817 from binary compact stars, we can now study the properties of compact stars through a completely new window, in particular using the observation limits on the tidal deformability as mentioned above. It will soon be possible to put constraints on those hypothetical phases within compact stars. 
	In \cite{2017PhRvD..95j1302L}, we propose that the tidal deformability of compact stars may give us a useful probe to solid quark stars due to the extreme rigidity of the CCS phase of quark matter.
	
	Penner {\it et al.} \cite{2011PhRvD..84j3006P} first studied the effect of elasticity on the tidal deformability of neutron stars using polytropic models with a thin elastic crust to mimic traditional neutron star models. They conclude that the elasticity of the neutron-star crust causes a tiny reduction of the tidal deformability compared to the fluid counterpart.
	On the other hand, our previous study \cite{2017PhRvD..95j1302L} reveals that the tidal deformability of a solid quark star can be up to about 60\% smaller than its fluid-star counterpart. This causes a significant deviation in the I-Love relation, which relates the moment of inertia ($I$) and the tidal deformability (sometimes quantified by the tidal Love number \cite{1911spge.book.....L}), of solid quark stars from the universal relation \cite{2013PhRvD..88b3009Y,2013Sci...341..365Y} discovered for fluid compact stars. (see \cite{2017PhR...681....1Y} for a review). As a result, the properties of solid quark stars containing the CCS quark matter can be constrained from the I-Love relation if the independent accurate measurements of $I$ \cite{2005ApJ...629..979L} and tidal deformability \cite{2008PhRvD..77b1502F}, $\lambda$, are available in the future.
	
	In this paper, we extend the study to three types of composite compact star models containing the CCS phase quark matter: (1) hybrid stars containing a solid CCS phase quark matter core and a fluid nuclear matter envelope \cite{2008PhRvD..77b3004I,2009PhRvD..79h3007K,2013PhRvD..88l4002L}, (2) dressed quark stars with a solid quark matter core in CCS phase and a thin nuclear matter crust separated by Coulomb force \cite{2015ApJ...815...81M}, (3) two-layer quark stars with a fluid CFL quark matter core and a solid CCS quark matter envelope \cite{2014PhRvD..89j3014M}. These models all contain two layers with distinct elastic properties. We focus on the effect on the tidal deformabilities brought by the rigid CCS phase and the influence on such an effect caused by a layer in different composition. For instance, we have found the so-called `screening effect' in \cite{2017PhRvD..95j1302L}, where the fluid envelope of a hybrid star masks the effect of elasticity of the CCS quark matter core on the tidal deformability so that the value of $\lambda$ for the hybrid star is essentially the same as that of a stellar model with a fluid quark matter core.
	
	As mentioned above, the calculation of the tidal deformability of traditional neutron stars with 
	an elastic crust in general relativity (GR) has been formulated by Penner {\it et al.} \cite{2011PhRvD..84j3006P}. We have formulated the problem ourselves and re-derived the set of equations using a different choice of variables comparing to \cite{2011PhRvD..84j3006P} so that the resulting matter equations can be compared directly to those corresponding Newtonian equations (e.g., \cite{2015Icar..248..109B,2015Icar..258..239B}) in elastic layers. We have applied our equations in the previous study \cite{2017PhRvD..95j1302L} and we present the full set of equations and the relevant boundary conditions in this paper.
	
	In Sec.~\ref{sec2:level1}, we present the formulation to compute the tidal deformability of two-layer compact stars with a solid component. 
	Section~\ref{sec4:level1} presents our numerical results for various two-layer compact star 
	models. In Sec.~\ref{sec:factors}, we study how the screening effect is affected by the 
	stellar structure and physical parameters. We also briefly discuss the relevance of our work to GW170817 in Sec.~\ref{sec:GW170817}. Finally, we conclude our paper in Sec.~\ref{sec:conclude}. Unless otherwise noted, we use geometric units with $G=c=1$. 
	
	%
	%
	%
	%
	%
	
	\section{\label{sec2:level1}Formulation}
	
	In this section, we shall present the full formulation of our calculations. Readers who are more interested in the physical results may skip this section and go to Sec.~\ref{sec4:level1} directly.
	
	The determination of the tidal deformation of a compact star requires full general relativistic treatment. We are interested in the weak field regime where the linear approximation is valid. Extensive studies have been done on the tidal deformability of compact stars with different configurations, including the static equilibrium models composed entirely of fluid \cite{2009PhRvD..80h4035D,2008ApJ...677.1216H,2009PhRvD..80h4018B,2010PhRvD..82b4016P,2010PhRvD..81l3016H} and slowly rotating fluid models \cite{2015PhRvD..91j4018L,2015PhRvD..92b4010P,2017PhRvD..95l4058L}. Our focus shall be on the tidal deformability of models with solid layers. The set of static perturbation equations in the solid crust was first derived by Penner~\textit{et~al.} \cite{2011PhRvD..84j3006P} from the Einstein field equations in their investigation of the tidal deformation of polytropic models with a solid crust. In fact, the static perturbation problem can also be considered as the zero frequency limit of the polar pulsation problem, which was first considered by Thorne and Campolattaro \cite{1967ApJ...149..591T} in relativistic fluid bodies and later formulated for solid bodies by Finn \cite{1990MNRAS.245...82F}. In this paper, we provide a new set of equations that is more suited for comparison with the Newtonian counterpart, allowing easy verifications and possible extension of the Newtonian analytical studies to relativistic cases.
	
	We derive the set of linearized static perturbation equations for polar deformations in compact stars to study the tidal deformation problem starting from the Einstein field equations and the continuity equations. We focus on developing a formalism in direct analogy to the conventional Newtonian perturbation equations for solid stellar models (see e.g., Alterman~\textit{et~al.} \cite{Alterman80}, Saito \cite{1974123}, Ushomirsky~\textit{et~al.} \cite{2000MNRAS.319..902U}) by choosing a specific set of dependent variables with Newtonian counterparts. We have also spotted some mistakes or typos in the perturbation equations of \cite{2011PhRvD..84j3006P} by comparing them with our equations and others' work (e.g., \cite{1990MNRAS.245...82F}).
	
	The determination of the tidal deformability consists of three steps. First, we find the equilibrium structure of the unperturbed compact star model with the Tolman-Oppenheimer-Volkov (TOV) equations and the EOS. Then, we solve the linearized static perturbation equations with the variables from background configuration. In this way, we obtain the linear response of the mass elements within the star under an arbitrary time-independent perturbation. Finally, we calculate the tidal deformability by solving for the external metric perturbation outside the star using the solutions of the perturbation problem in the stellar interior.
	
	\subsection{\label{sec2:level2A}Equilibrium background}
	
	The spacetime metric for a spherically symmetric, static equilibrium background is given by 
	\begin{equation} \label{sec2:eqt1}
	ds^2 = - e^{\nu(r)} dt^2 + e^{\lambda(r)} dr^2 + r^2 (d\theta^2 + \sin^2\theta d\phi^2).
	\end{equation}
	The structure of a compact star is governed by the TOV equations
	\begin{align}
	\frac{d\nu(r)}{dr} &= \frac{2 e^{\lambda(r)}}{r^2}\Big(m(r) + 4 \pi r^3 P(r)\Big), \label{sec2:eqt2} \\ 
	\frac{dP(r)}{dr} &= -\frac{\rho(r) + P(r)}{2}\nu^\prime(r), \label{sec2:eqt3} \\ 
	\frac{dm(r)}{dr} &= 4 \pi r^2 \rho(r),  \label{sec2:eqt4}
	\end{align}
	where $m$ is the gravitational mass within a radius $r$ and the functions $P$, $\rho$ are the pressure, energy density of a mass element at a distance $r$ from the center respectively. In this context, we also use the prime symbol to denote radial derivative. For instance, $\nu^\prime(r)$ represents $d\nu(r) / dr$ in Eq.~(\ref{sec2:eqt3}). For a cold compact star, the zero temperature EOS takes a simple form: $P = P(\rho)$. With this additional information given, we solve Eqs.~(\ref{sec2:eqt2})-(\ref{sec2:eqt4}) to determine the equilibrium stellar structure. The function $\lambda$ is given by
	\begin{equation} \label{sec2:eqt4_5}
	e^{\lambda(r)} = \frac{1}{1 - 2m(r)/r}.
	\end{equation}
	
	\subsection{\label{sec2:level2B}Static perturbation equations}
	
	To calculate the tidal deformability, we must first obtain the solution to the relativistic static perturbation problem of the stellar interior with appropriate boundary conditions. Assume that the star deforms under an external tidal field, which induces a mass quadrupole moment inside the star. The perturbations are governed by the linearized Einstein field equations and the continuity equations
	\begin{equation}\label{sec2:eqt5}
	\delta G_{\alpha \beta} = 8 \pi \delta T_{\alpha \beta},
	\end{equation}
	\begin{equation}\label{sec2:eqt6}
	\delta \big(T^{\alpha \beta}_{\quad ; \alpha}\big) = 0, 
	\end{equation}
	where $G_{\alpha \beta}$ is the Einstein tensor and $T_{\alpha \beta}$ is the stress-energy tensor. The semi-colon represents covariant derivatives. Unless specified otherwise, we use `$\delta$' to denote Eulerian perturbations.
	
	The perturbations, decomposed in the basis of spherical harmonics, can be classified into axial modes and polar modes based on their parities (\cite{1957PhRv..108.1063R,1967ApJ...149..591T}). We focus on the even parity perturbations in our tidal deformation problem.
	\subsubsection{Fluid perturbation equations}
	
	Considering polar perturbations of a static spherically symmetric background metric in the Regge-Wheeler gauge \cite{1957PhRv..108.1063R}, the metric perturbation of frequency $\omega$ is expressed as  
	
	\begin{equation}
	\delta g_{ab} = h_{ab}(r)Y_{lm}(\theta, \phi) e^{i\omega t},
	\end{equation}
	where
	\begin{equation}\label{Metric_Perturbed}
	h_{ab}(r) = 
	\left(
	\begin{array}{cccc}
	H_0(r) e^\nu & i \omega H_1(r) & 0 & 0 \\ 
	i \omega H_1(r) & H_2(r) e^\lambda & 0 & 0 \\ 
	0 & 0 & r^2 K(r) & 0 \\ 
	0 & 0 & 0 & r^2  \sin^2\theta K(r)
	\end{array} 
	\right).
	\end{equation}
	The displacement vector of polar perturbation is given by
	\begin{align}
	\xi^r &= \frac{W(r)}{r}~Y_{lm}(\theta, \phi), \\
	\xi^\theta &= \frac{V(r)}{r^2}~\partial_\theta Y_{lm}(\theta, \phi), \\
	\xi^\phi &= \frac{V(r)}{r^2 \sin^2\theta}~\partial_\phi Y_{lm}(\theta, \phi),
	\end{align}
	where $Y_{lm}(\theta, \phi)$ is the standard spherical harmonics function. For static perturbations, we set $\omega = 0$. This leaves only the diagonal terms non-zero in the perturbed metric. In perfect fluid, the perturbed stress-energy tensor is written in terms of the energy density $\rho$, pressure $P$, four-velocity $U^\alpha$ and their corresponding perturbed quantities 
	\begin{equation}\label{Stress_Tensor}
	\begin{split}
	\delta T^{\alpha \; \text{fluid}}_{\;\beta} =& (\delta \rho + \delta P) U^\alpha U_\beta + \delta P \; \delta^\alpha _{\;\beta} \\ 
	&+ (P + \rho) (U^\alpha \delta U_\beta + \delta U_\alpha U^\beta ),
	\end{split}
	\end{equation}
	where $\delta^\alpha _{\;\beta}$ represents the Kronecker delta function.
	From the linearized Einstein field equations, the perfect fluid problem is cast into a single second order differential equation of $H_0$ \cite{2008ApJ...677.1216H}:
	\begin{equation}\label{Fluid_Perturbation_01}
	\begin{split}
	&\frac{d^2 H_0}{dr^2} +\frac{d H_0}{d r} \bigg[\frac{2}{r} + e^{\lambda} \Big(\frac{2m(r)}{r^2} + 4 \pi r (P - \rho)\Big)\bigg] \\ 
	&+ H_0 \left[-\frac{l (l + 1) e^{\lambda}}{r^2} + 4\pi e^{\lambda} \left(5 \rho + 9 P + \frac{\rho + P}{{c_s}^2}\right) - {\nu^\prime}^2\right] = 0,
	\end{split}
	\end{equation}
	where ${c_s}^2 = dP/d\rho$.
	
	\subsubsection{Solid perturbation equations}
	
	To account for elasticity, we first assume the background to be in an unstrained state, given that the background shear only affects the total stress energy by a negligible amount. In this way shear only contributes in the perturbation level. Hence, the total stress energy tensor in Eqs.~(\ref{sec2:eqt5}) and (\ref{sec2:eqt6}) is written as \cite{1990MNRAS.245...82F,2011PhRvD..84j3006P}
	\begin{equation}
	\delta T_ {\alpha \beta} = \delta T^{\text{bulk}} _ {\alpha \beta} + \delta T^{\text{shear}}_{\alpha \beta},
	\end{equation}
	where the effect of shear is given in the anisotropic stress tensor $\delta T^\text{shear}_{\alpha \beta}$ following a Hookean relationship with the shear strain tensor $\delta \Sigma_{\alpha \beta}$ and shear modulus $\mu$
	\begin{equation}
	\delta T^{\text{shear}}_{\alpha \beta} = -2 \mu \; \delta \Sigma_{\alpha \beta}.
	\label{Hookes_Law}
	\end{equation}
	Following \cite{1972RSPSA.331...57C}, the shear strain tensor for small deformations obeys the differential equation
	\begin{equation} \label{Strain_Rate}
	\frac{d \delta \Sigma_{\alpha \beta}}{d \tau} = \frac{1}{2}\Big( \perp^\mu_{\; \alpha} U_{\beta ; \mu} +  \perp^\mu_{\; \beta} U_{\alpha ; \mu}\Big) - \frac{1}{3} \perp^{\mu \nu} \perp_{\alpha \beta}U_{\mu ; \nu} ,
	\end{equation}
	with the projection tensor $\perp^\mu_{\; \nu}$ defined by
	\begin{equation}
	\perp^\mu_{\; \nu} = \delta^\mu_{\; \nu} + U^\mu U_\nu.
	\end{equation}
	Eq.~(\ref{Strain_Rate}) is solved to the linear order by Finn \cite{1990MNRAS.245...82F}. Directly applying the results in \cite{1990MNRAS.245...82F}, we write down the strain tensor components represented by the radial and tangential strain variables, $S_r$ and $S_\perp$ respectively, which are defined similarly to those in \cite{1990MNRAS.245...82F}:
	\begin{align}
	&\delta \Sigma^r_r \nonumber\\
	\equiv&  S_r(r) \; Y_{lm}(\theta , \phi) \nonumber\\
	=& \frac{1}{3} \left\{H_2-K + \frac{l(l+1)}{r^2}V + \frac{2}{r} \frac{dW}{dr} - \left(\frac{4}{r^2} - \frac{\lambda^{\prime}}{r}\right)W\right\} Y_{lm},\nonumber\\  \label{strain_r}
	\end{align}
	\begin{align} \label{strain_t}
	\frac{1}{r}\delta \Sigma^{r}_{A} \equiv& S_{\perp}(r) \; \partial_A Y_{lm}(\theta , \phi) \nonumber \\
	=& \frac{e^{-\lambda}}{2 r} \left\{\frac{dV}{dr} - \frac{2V}{r} + e^{\lambda}\frac{W}{r}\right\}\;  \partial_A Y_{lm}.
	\end{align}
	The remaining components are
	\begin{equation}
	\delta \Sigma^{A}_{B} = V(r) Y_{lm}^{\; ,A} \, _{;B} - \frac{1}{2} \Big[S_r(r) - \frac{l(l+1)}{r^2} V(r) \Big] \delta^A_{\; B} Y_{lm},
	\end{equation}
	where the indexes $A$ and $B$ both run over $\theta$ and $\phi$. The comma signs before the indexes denote partial derivatives. The shear strain tensor is guaranteed to be traceless in the above expressions. The contributions from shear are determined from the three radial functions $S_r(r)$, $S_\perp (r)$ and $V(r)$ and the shear modulus $\mu$.
	
	Putting the above metric perturbations and stress energy perturbations into Eqs.~(\ref{sec2:eqt5}) and (\ref{sec2:eqt6}), we obtain the differential equations governing the solid polar perturbation problem.
	
	We define new variables $Z_r$ and $Z_\perp$ to represent the radial components of the total stress in radial and tangential directions respectively:
	\begin{equation}\label{Z_r}
	Z_r(r) = \Delta P(r) - 2 \mu S_r(r),
	\end{equation}
	\begin{equation}\label{Z_t}
	Z_\perp(r) = -2 \mu S_\perp(r),
	\end{equation}
	where $\Delta P$ is the Lagrangian perturbation of pressure. We also define a variable $J$:
	\begin{equation} \label{eq_def_J}
	J = H_0^\prime - 8\pi e^\lambda (\rho + P) \frac{W}{r} + 16 \pi \nu^\prime \mu V.
	\end{equation}
	The complete set of perturbation equations is given by
	\begin{widetext}
		\begin{align}
		\frac{d W}{dr} =& \left(1 - \frac{2 \alpha_2}{\alpha_3} - \frac{r \lambda^\prime}{2}\right)\frac{W}{r} - \frac{r}{\alpha_3} Z_r + \frac{\alpha_2}{\alpha_3}L_1 \frac{V}{r} - \frac{\alpha_2}{\alpha_3} r K - \frac{1}{2} r H_2, \label{fn_eqt1} \\
		\nonumber \\
		\frac{d Z_r}{dr} =& \left[ P^\prime \left( \frac{r \nu^{\prime \prime}}{\nu^\prime} - \frac{r \lambda^\prime}{2} - 2 \right) - \frac{4}{r} \frac{\alpha_1 }{\alpha_3} \left(\alpha_3 + 2 \alpha_2 \right)  \right] \frac{W}{r^2} - \left( \frac{r \nu^\prime}{2} + \frac{4 \alpha_1}{\alpha_3}\right) \frac{Z_r}{r}  \nonumber \\
		&+\left[P^\prime + \frac{2}{r}\frac{\alpha_1}{\alpha_3}\Big(\alpha_3 + 2 \alpha_2\Big) \right]\frac{L_1 V}{r^2}+ \frac{e^\lambda L_1}{r} Z_\perp + \frac{1}{2}(\rho + P) H_0^\prime  - \frac{P^\prime}{2}H_2 \nonumber \\
		&- \left[P^\prime + \frac{2}{r} \frac{\alpha_1 }{\alpha_3} \left(\alpha_3 + 2 \alpha_2 \right) \right]K, \label{fn_eqt2}\\
		\nonumber \\
		\frac{d V}{dr} =& -e^\lambda \frac{W}{r} + \frac{2 V}{r} -\frac{r e^\lambda}{\alpha_1} Z_\perp, \label{fn_eqt3}\\
		\nonumber \\
		\frac{d Z_\perp}{dr} =& \left[P^\prime + \frac{2}{r}\frac{\alpha_1}{\alpha_3}\left(\alpha_3 + 2 \alpha_2\right)\right]\frac{W}{r^2} - \frac{\alpha_2}{\alpha_3} \frac{Z_r}{r}- \left[-\frac{2 \alpha_1}{r} + 2 \alpha_1 \left(1 + \frac{\alpha_2}{\alpha_3}\right)\frac{L_1}{r}\right] \frac{V}{r^2} \nonumber\\
		& - \left[\frac{r \lambda^\prime}{2}+ \frac{r \nu^\prime}{2} + 3\right] \frac{Z_\perp}{r}  +\frac{1}{2}\left(\rho + P\right) \frac{H_0}{r}+ \frac{\alpha_1}{\alpha_3} \left(\alpha_3 + 2 \alpha_2 \right) \frac{K}{r}, \label{fn_eqt4}\\
		\nonumber \\
		\frac{d H_0}{d r} =& J + \frac{1}{r^2}(\nu^\prime + \lambda^\prime) W - 16 \pi \alpha_1 \nu^\prime V, \label{fn_eqt5}\\
		\nonumber \\
		\frac{d J}{d r} =& \left[\frac{32 \pi e^\lambda}{r^2} \frac{ \alpha_1}{\alpha_3}\left(\alpha_3 + 2 \alpha_2\right) -\frac{3}{2 r^2}\nu^\prime\left(\lambda^\prime + \nu^\prime\right)\right] W- 8 \pi e^\lambda \frac{1}{\alpha_3}\left( \alpha_3 + 2 \alpha_2\right)Z_r \nonumber\\
		&  -\frac{8 \pi}{r^2}\Big[\big(\rho + P \big)e^\lambda L_1+ \frac{2 \alpha_1}{\alpha_3}\big(\alpha_3 + 2 \alpha_2\big) e^\lambda L_1 + 4 \alpha_1 \big(1 - e^\lambda\big) - 2\alpha_1 \big(r \nu^\prime\big)^2\Big] V \nonumber\\
		& - 16 \pi e^\lambda \big(r \nu^\prime\big)Z_\perp + \left[L_1 e^\lambda + 2\big(e^\lambda - 1\big) - r\left(\frac{\lambda^\prime}{2} + \frac{5 \nu^\prime}{2}\right) + \big(r \nu^\prime\big)^2\right]\frac{H_0}{r^2} \nonumber\\ 
		&+ \left[\frac{r}{2}\big(\lambda^\prime - \nu^\prime \big) - 2\right] \frac{J}{r} +  \left[\frac{1}{r}\big(\lambda^\prime + \nu^\prime\big) + 16 \pi e^\lambda \frac{\alpha_1}{\alpha_3}\big(\alpha_3 + 2\alpha_2\big)\right]K, \label{fn_eqt6}
		\end{align}
	\end{widetext}
	where
	\begin{align}
	L_1 = l(l+1),
	\end{align}
	and
	\begin{align} \label{fn_H2}
	H_2 &= H_0 + 32 \pi \alpha_1 V.
	\end{align}
	In the above equations, we define a set of quantities similar to those in McDermott~\textit{et~al.}~\cite{1988ApJ...325..725M}, which studies the non-radial pulsations in neutron stars with a solid crust with Newtonian Cowling approximation, to represent different elastic moduli in isotropic materials:
	\begin{equation}
	\begin{cases} 
	&\alpha_1= \mu, \\
	&\alpha_2= {c_s}^2 (\rho + P) - \frac{2}{3}\mu, \\ 
	&\alpha_3= {c_s}^2 (\rho + P) + \frac{4}{3}\mu, \\
	\end{cases}
	\end{equation}
	where $\alpha_2$ and $\alpha_3$ represent the relativistic generalization of the Lam\'e coefficient and the P-wave modulus defined in classical elastic theory respectively (see e.g., \cite{2003rph..book.....M}). The metric perturbation variable $K$ is expressed in terms of the other perturbation variables by
	\begin{equation} \label{fn_alg1}
	\begin{split}
	&(L_1 -2) \; e^\lambda K = \left[\big(\nu^\prime\big)^2 + \nu^\prime \lambda^\prime +16 \pi e^\lambda r P^\prime\right] W \\
	& - 16\pi e^\lambda r^2 Z_r - 16 \pi e^\lambda \left(2 + r \nu^\prime \right) r^2 Z_\perp  \\
	&+ \left[L_1 e^\lambda - 2+ \big(r \nu^\prime \big)^2 \right] H_0 + r^2 \nu^\prime J.
	\end{split}
	\end{equation}
	
	Eqs.~(\ref{fn_eqt1})-(\ref{fn_eqt4}) can be derived solely from the continuity equation (Eq.~(\ref{sec2:eqt6})). They reproduce the zero-frequency limit of the equations governing the polar pulsation in solid under relativistic Cowling approximation given in \cite{2002A&A...395..201Y} if we neglect the metric perturbations. Meanwhile, Eqs.~(\ref{fn_eqt5}) and (\ref{fn_eqt6}) are obtained from the perturbed Einstein field equations (Eq.~(\ref{sec2:eqt5})).
	
	Eqs.~(\ref{fn_eqt1})-(\ref{fn_eqt6}), together with the algebraic relations for $H_2$ and $K$, Eqs.~(\ref{fn_H2}) and (\ref{fn_alg1}), form a complete set of perturbation equations in solid with variables $\big(W$, $Z_r$, $V$, $Z_\perp$, $H_0$, $J\big)$. We have checked that our equations are consistent with the zero frequency limits of the two sets of relativistic non-radial pulsation equations for polar modes in solid compact stars given independently by Finn \cite{1990MNRAS.245...82F} and Kr{\"u}ger \textit{et al.} \cite{2015PhRvD..92f3009K} (See the footnote \footnote{Note that \cite{2015PhRvD..92f3009K} contains a minor typo in all the terms containing the shear modulus in the pulsation equations which makes them two times larger than the actual terms. The same typo is also found in \cite{2011PhRvD..84j3006P}.}). We also notice that the equations for static perturbations in solid by Penner~\textit{et~al.} \cite{2011PhRvD..84j3006P} are inconsistent with the zero frequency limits of the above mentioned pulsation equations in \cite{1990MNRAS.245...82F} and \cite{2015PhRvD..92f3009K}.
	
	\subsection{\label{sec2:level2C}Boundary conditions}
	
	\subsubsection{Conditions at stellar center}
	
	The set of perturbation equations in solid has a regular singular point at the origin. We derive the regular solutions of the perturbation equations for solid core near the origin by expanding the six perturbation variables about $r=0$. The leading power dependence of these variables are found by solving the indicial equations and the results are given in Appendix~\ref{app1}. Keeping the first two non-zero terms of the expansion of each variable, which gives a total of twelve coefficients (Eq.~(\ref{app1_expansion})), we find nine independent constraints (Eqs.~(\ref{static_origin_regular_01})-(\ref{static_origin_regular_09})) by substituting the series expansions into the system of six first-order differential equations (Eqs.~(\ref{fn_eqt1})-(\ref{fn_eqt6})). This gives three independent regular solutions at the origin. For stellar models with a fluid core, we refer the readers to previous works (e.g., \cite{2008ApJ...677.1216H}) for the corresponding boundary conditions.
	
	\subsubsection{Conditions at interface and stellar surface}
	
	Across the solid-fluid interface, the shear modulus exhibits a jump from a finite value in the solid layer to zero in the layer of perfect fluid. Furthermore, the phase transition from quark matter to nuclear matter in hybrid star models determined by the Maxwell construction has a density discontinuity. As a result, some of the perturbation variables are not continuous across the interfaces. Boundary conditions are imposed to relate the perturbation variables at the two sides of the interfaces.
	
	The continuity of intrinsic curvature required by the Einstein field equations (see Finn \cite{1990MNRAS.245...82F} for detailed derivation) implies that the variables $H_0, K, W$ must be continuous across the interface. The fact that the stress energy tensor being nonsingular at any point leads to the continuity of $Z_r$ and $Z_\perp$ across the perturbed interface \cite{2011PhRvD..84j3006P} (i.e., Israel junction condition \cite{1966NCimB..44....1I}). From the above continuity conditions, we clearly see the advantage of using the function $J$ (see Eq.~(\ref{eq_def_J})) as the dependent variable over using $H_0^\prime$, as $J$ is continuous across the interface and the stellar surface while $H_0^\prime$ is not.
	
	In our formulation, we apply the continuities of  $\left(W, Z_r, Z_\perp, H_0, J\right)$ across the solid-fluid interface. In the statically perturbed fluid layer, the displacement variables $W$ and $V$ are undetermined except at the interface and surface. Therefore, we express the radial stress variable $Z_r$ at the fluid side of the interface in terms of $H_0$ and $W$. With the algebraic relation for $H_0$ and $\delta P$ in fluid, 
	\begin{equation}
	\delta P = \frac{1}{2}(\rho + P) H_0,
	\end{equation}
	we explicitly give the continuities of $Z_r$ and $Z_\perp$ as
	\begin{equation} \label{interface_eqt3}
	\frac{1}{2}\Big(\rho^{(\text{f})} + P\Big) \bigg(H_0 - \nu^\prime \frac{W}{r}\bigg) = Z_r^{(\text{s})},
	\end{equation}
	\begin{equation} \label{interface_eqt4}
	Z_\perp^{(\text{f})} = Z_\perp^{(\text{s})} = 0,
	\end{equation}
	where the continuity conditions are imposed at the interface. We use the superscripts `$(\text{f})$' and `$(\text{s})$' on the quantities that are in general discontinuous to indicate the fluid side and the solid side of the interface respectively. Eqs.~(\ref{interface_eqt3}) and (\ref{interface_eqt4}) allow us to determine the solution in the solid core up to an arbitrary constant. The remaining part in the fluid envelope is an initial value problem from the interface to the stellar surface.
	
	The stellar surface can be treated as an interface between the star interior and vacuum. The relevant boundary conditions are similar to those at the solid-fluid interface. For a solid-vacuum interface, i.e., at the stellar surface of bare solid quark stars, the boundary conditions are written explicitly as
	\begin{align}
	H_0(R_+) &= H_0(R_-), \label{surface_eqt1}\\
	H_0^\prime(R_+) &= J(R_+) = J(R_-), \label{surface_eqt2}\\
	Z_r(R_+) &= Z_r(R_-) = 0, \label{surface_eqt3}\\
	Z_\perp(R_+) &= Z_\perp(R_-) = 0, \label{surface_eqt4}
	\end{align}
	where $R$ is the stellar surface, the plus and minus signs in the subscripts of $R$ indicate the outer side and inner side of the stellar surface respectively. In the bare quark star models, there are three independent regular solutions of unknown amplitudes in the star interior. The amplitude of each of the solutions is fixed up to an arbitrary constant by Eqs.~(\ref{surface_eqt3}) and (\ref{surface_eqt4}). After determining the interior solution, Eqs.~(\ref{surface_eqt1}) and (\ref{surface_eqt2}) are used to determine the metric perturbation in the vacuum side of the stellar surface, which allows us to calculate the tidal deformability. 
	
	Note that only Eqs.~(\ref{surface_eqt1}) and  (\ref{surface_eqt2}) are relevant across a fluid-vacuum interface in determining the tidal deformability as $Z_\perp$ is always zero in fluid (perfect fluid assumption) and Eq.~(\ref{surface_eqt3}) provides an extra relation between $W(R_-)$ and $H_0(R_-)$:
	\begin{equation}
	Z_r(R_-) = \frac{1}{2}\rho H_0 + P^\prime \frac{W}{r} \Big|_{R_-} = 0.
	\end{equation}
	
	\subsection{\label{sec2:level2D}Tidal deformability}
	
	After obtaining the values of the metric perturbation at the vacuum side of the stellar surface, the tidal Love number is calculated using the same method for a fluid star. The following gives a brief description on the procedure. We refer the readers to \cite{1967ApJ...150.1005H,1968ApJ...153..807H,2008ApJ...677.1216H,2009PhRvD..80h4035D} for more detailed discussions.
	
	The metric for a static, spherically symmetric stellar model under a static external tidal field in the far-field limit is given by Thorne \cite{1998PhRvD..58l4031T}:
	\begin{align}
	\label{Vacuum_Love_metric}
	-\frac{1 + g_{tt}}{2} =& -\frac{M}{r} - \frac{3Q_{ij}}{2 r^3} \left(\frac{x^i x^j}{r^2} - \frac{1}{3}\delta^{ij}\right) \\ 
	&+ \frac{1}{2}\mathcal{E}_{ij}x^i x^j, \nonumber 
	\end{align}
	where $Q_{ij}$ is the quadrupole moment and $\mathcal{E}_{ij}$ is the external tidal field. The tidal Love number, $k_2$, of a static spherically symmetric star is defined by the relation \cite{2008ApJ...677.1216H}
	\begin{equation}
	\label{Love_Def}
	Q_{ij} = -\frac{2}{3}k_2 R^5 \mathcal{E}_{ij},
	\end{equation}
	where $R$ is the star radius.
	
	The behavior of $H_0$ in vacuum for the perturbed system is governed by the equation \cite{2008ApJ...677.1216H}:
	\begin{equation}\label{Vacuum_Love}
	H_0^{\prime \prime} + \left(\frac{2}{r} - \lambda^\prime \right) H_0^\prime - \left[\frac{l(l+1)e^{\lambda}}{r^2} - \left(\lambda^\prime\right)^2\right]H_0 = 0.
	\end{equation}
	Using the change of variables $x = r/M - 1$ as in  \cite{1967ApJ...149..591T,2008ApJ...677.1216H}, Eq.~(\ref{Vacuum_Love}) is transformed into the standard associated Legendre equation of order $(l, m)$ with $m = 2$. The solutions are the associated Legendre functions $Q_l^2(x)$ and $P_l^2(x)$
	
	\begin{equation}\label{Vacuum_Love_Soln}
	H_0(r) = c_1 Q_l^2(x) + c_2 P_l^2(x).
	\end{equation}
	The asymptotic behavior of the associated Legendre functions are
	\begin{align}
	Q_l^2(x) &\sim \left(\frac{M}{r}\right)^{l+1},\\
	P_l^2(x) &\sim \left(\frac{r}{M}\right)^{l}, 
	\end{align}
	respectively. Using Eqs.~(\ref{Vacuum_Love_metric}) and (\ref{Love_Def}) in the far-field limit to determine the coefficients in Eq.~(\ref{Vacuum_Love_Soln}), we can obtain the Love number $k_2$ by
	\begin{equation}\label{Love_Exterior}
	k_2 = \frac{4}{15}\left(\frac{M}{R}\right)^5 \frac{c_1}{c_2}.
	\end{equation}
	The value of $c_1 / c_2$ depends on the solution of $H_0$ and its derivative at the vacuum side of the stellar surface. Using the interior solution and the relevant boundary conditions Eqs.~(\ref{surface_eqt1}) and (\ref{surface_eqt2}), we can express Eq.~(\ref{Love_Exterior}) in terms of the compactness $C = M/R$ and the dimensionless parameter $y$ \cite{2008ApJ...677.1216H}:
	\begin{widetext}
		\begin{align}
		k_2 =& \left\{ \frac{8C^5}{5}(1-2C)^2 \left[2 + 2C\left(y - 1\right) - y\right]\right\} \Bigg\{2C\left[4(1+y)C^4 + (6y-4)C^3 + (26-22y)C^2 + 3C(5y-8)-3y+6 \right] \nonumber \\
		&+ 3(1-2C)^2\left[2 - y + 2C (y - 1)\right] \text{log}(1 - 2C)\Bigg\}^{-1},
		\end{align}
	\end{widetext}
	where $y$ is defined by the values of $H_0$ and its derivative at the vacuum side of the stellar surface:
	\begin{equation}
	y = r\frac{H_0^\prime(r)}{H_0(r)}\bigg\vert_{r=R_+}.
	\end{equation}
	
	In the following discussion, we mainly quantify the tidal deformability with the quantity, $\bar{\lambda}$, namely the `normalized tidal deformability', defined by 
	\begin{equation} \label{tidal_deformability}
	\bar{\lambda} = \frac{2}{3}k_2 \left(\frac{R}{M}\right)^{5},
	\end{equation}
	which appears in the discussion of the universal I-Love-Q relations in fluid compact star models \cite{2013PhRvD..88b3009Y, 2013Sci...341..365Y}.
	
	%
	%
	%
	%
	%
	%
	
	%
	%
	%
	%
	%
	
	\section{\label{sec4:level1}Numerical Results}
	
	In \cite{2017PhRvD..95j1302L}, we report that the deviation of the tidal deformability of solid quark stars composed entirely of CCS phase quark matter from a fluid quark star with the same background profile can potentially be as large as 60\% in $\lambda$. In the following, we shall study the tidal deformability of three different types of composite models and compare the results with that of solid quark stars.
	
	\subsection{Hybrid stars with a solid quark matter core and a fluid nuclear matter envelope} \label{ssec:BG}
	
	In this study, we describe quark matter with the phenomenological model proposed by Alford \textit{et al.} \cite{2005ApJ...629..969A}. The EOS is given by the grand potential per unit volume, $\Omega_\text{QM}$:
	\begin{equation} \label{QM_EOS}
	\Omega_\text{QM} = -\frac{3}{4\pi^2} a_4 \mu_{q}^4 + \frac{3}{4\pi^2}a_2 \mu_q^2 + B_{\text{eff}},
	\end{equation}
	where $\mu_{q}$ is the average quark chemical potential of the mixture of up, down and strange quarks. $a_4~(\leq 1)$ is a parameter used to model the non-perturbative QCD corrections, with a typical value of around 0.7 (see \cite{2001PhRvD..63l1702F}). $a_2$ depends on both the strange quark mass and color superconducting gap to take account of the free energy correction due to quark masses and quark pairing. $B_{\text{eff}}$ is the effective bag constant related to the vacuum pressure. A hybrid star consists of a nuclear matter envelope in addition to the quark matter core. We choose the EOS model APR \cite{1998PhRvC..58.1804A} to describe the nuclear matter. The phase transition between the core and the envelope is determined with the Maxwell construction (see \cite{2005ApJ...629..969A,2013PhRvD..88l4002L}). The resulting model has a finite density gap at the core-envelope interface. On the other hand, a bare quark star contains quark matter only and is described by the phenomenological EOS.

	Besides the EOS, we also need the shear modulus $\mu$ in our calculations. The shear modulus of the CCS quark matter is given approximately by \cite{2007PhRvD..76g4026M}
	\begin{equation}
	\mu = 2.47 \; \text{MeV fm}^{-3} \bigg(\frac{\Delta}{10~\text{MeV}}\bigg)^2 \bigg(\frac{\mu_q}{400~\text{MeV}}\bigg)^2,
	\end{equation}
	where the value of the gap parameter $\Delta$ is expected to lie within the range 5~MeV to 25~MeV.
	
	An interesting phenomenon found in our previous study \cite{2017PhRvD..95j1302L} is that a fluid envelope is able to strongly screen out the effect of elasticity of the solid core if the density gap at the core-envelope interface is small compared to the core density. In particular, the tidal deformability of a solid quark star can deviate from that of a fluid quark star to 60\%, while the difference between a hybrid star with a CCS solid core and that with a fluid core is only around 1\% for a particular hybrid star model investigated in \cite{2017PhRvD..95j1302L}. This demonstrates that even though the elastic CCS quark matter can have a significant impact on the tidal deformability of bare quark stars, a fluid envelope on the surface may cancel out this effect. In this section, we further investigate the screening effect in hybrid stars with different EOS models to understand how such an effect can affect hybrid stars with different internal structures.
	
\begin{figure}[!t]
		\centering
		\includegraphics[width=8.6cm]{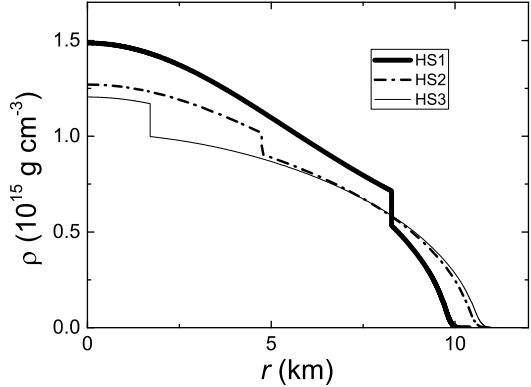}
		\caption{\label{Density_HS_14} The density profiles of the hybrid star models of 1.4~$M_\odot$. All the models have a density gap of around 20\% of the central densities at the core-envelope interface.}
	\end{figure}

	Using the hybrid star EOS described above, with the parameters listed in Table~\ref{EOS_parameter} for the quark matter EOS Eq.~(\ref{QM_EOS}), we construct three hybrid star models (HS1 - HS3) of 1.4~$M_\odot$ with a solid core and compare their normalized tidal deformabilities, which are first constructed in \cite{2013PhRvD..88l4002L} in the study of torsional oscillations of hybrid stars with CCS quark matter. For comparison, we also consider a solid bare quark star model, abbreviated as `SQS' in the following, constructed with the same set of EOS parameters as that of the quark matter EOS for the core of HS2. The mass of SQS is also 1.4~$M_\odot$. The density profiles of the three hybrid star models are given in Fig.~\ref{Density_HS_14}. The models have similar radii but HS1 has the largest quark matter core while HS3 has the smallest. This allows us to qualitatively study the influence of the size of the CCS core on the screening effect.
	
	\begin{table}[!b]
		\centering
		\begin{tabular}{rccc|cc}
			\hline \hline
			\multicolumn{1}{c}{EOS} & \multicolumn{1}{c}{$a_4$} & \multicolumn{1}{c}{$a_2^{1/2}$~(MeV)} & \multicolumn{1}{c}{$B_{\text{eff}}^{1/4}$~(MeV)} &
			\multicolumn{1}{|c}{${M_{\text{e}}}/{M}$}&
			\multicolumn{1}{r}{$\bar{\lambda}_{\Delta  =25~\text{MeV}}$}\rule{0pt}{3ex} \\ \hline
			HS1 & 0.85 & 100 & 160 & 0.190 & 106.354\\
			HS2 & 0.8 & 100 & 160 & 0.749 & 206.504\\
			HS3 & 0.9 & 200 & 150 & 0.991 & 243.856\\
			\hline
		\end{tabular}%
		\caption{The parameters of the phenomenological EOS model for quark matter inside the core of the hybrid star models. The EOS of the nuclear matter envelope is APR \cite{1998PhRvC..58.1804A} for all three models. The last two columns refer to the parameters of the corresponding hybrid star models with 1.4~$M_\odot$, adapted from \cite{2013PhRvD..88l4002L}. $M_{\text{e}}$ and $M$ are the mass of the fluid envelope and the total mass of the hybrid star respectively. $\bar{\lambda}_{\Delta  =25~\text{MeV}}$ denotes the normalized tidal deformabilities of the hybrid stars with the gap parameter $\Delta = 25~\text{MeV}$.}
		\label{EOS_parameter}%
	\end{table}%
	
\begin{figure}[!t]
		\centering
		\includegraphics[width=8.6cm]{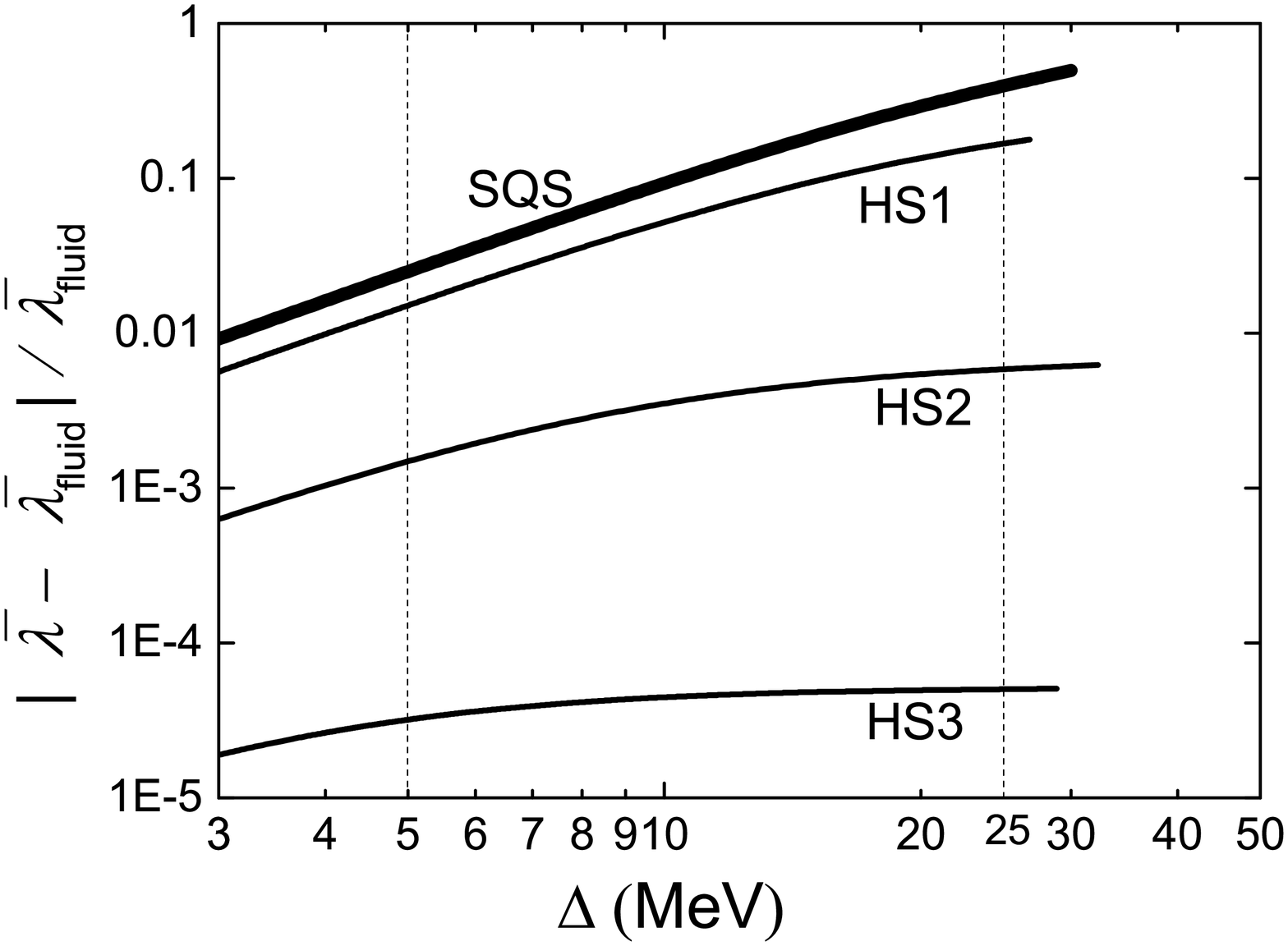}
		\caption{The fractional deviation in normalized tidal deformability against the gap parameter, $\Delta$, of hybrid stars HS1, HS2, HS3 and solid quark star SQS. 
			The parameter, $\bar{\lambda}_{\textrm{fluid}}$, here is the normalized tidal deformability of the fluid counterpart of that model with the same EOS but composed entirely of fluid as defined in the main text.
			The theoretical range of $\Delta$ is bounded by two vertical dashed lines.}	\label{lo:fig2}
	\end{figure}

	Figure~\ref{lo:fig2} compares the fractional deviation of the normalized tidal deformability, ${\Big| \bar{\lambda} - \bar{\lambda}_{\textrm{fluid}}\Big|}/{\bar{\lambda}_{\textrm{fluid}}}$, of the bare solid quark star and the three hybrid star models with different gap parameters $\Delta$ for the CCS phase (i.e., different shear moduli), where $\bar{\lambda}_{\textrm{fluid}}$ is the normalized tidal deformability of the corresponding fluid model with the same EOS but a zero shear modulus. This illustrates the large difference in the effect of elasticity on the tidal deformability between a solid quark star and hybrid stars.
	Figure~\ref{lo:fig2} also shows that the fractional deviation for hybrid stars decreases drastically by orders of magnitude as the thickness (as well as the mass) of the fluid envelope increases (from HS1 to HS3), which indicates a stronger screening effect as one might expect Our result show that the screening effect in our hybrid star models is very strong, with the fractional deviation smaller than the 0.01 level, as long as the CCS core size is smaller than about 70\% of the stellar radius.
	
	It is noted that the fractional deviation of HS1 is about 50\% of that of SQS. One might naively expect it to reduce to similar values as that of the SQS model if the thickness of the fluid envelope is further reduced. However, this is not the case. In fact, we find that the screening effect does not necessarily vanish even when the thickness (hence the mass) of the fluid envelope approaches zero. Showing this phenomenon with the hybrid star EOS is not easy as the core-envelope transition cannot be freely adjusted while keeping the other parameters, like the mass and the radius, unchanged. In particular, the core-envelope transition and hence the thickness of the envelope are fixed by the Maxwell construction. Therefore, we shall carry out further studies with some `toy models' that allow us to adjust the structures in Sec.~\ref{sec:factors}.
	
	\begin{figure}[!t]
		\centering
		\includegraphics[width=8.6cm]{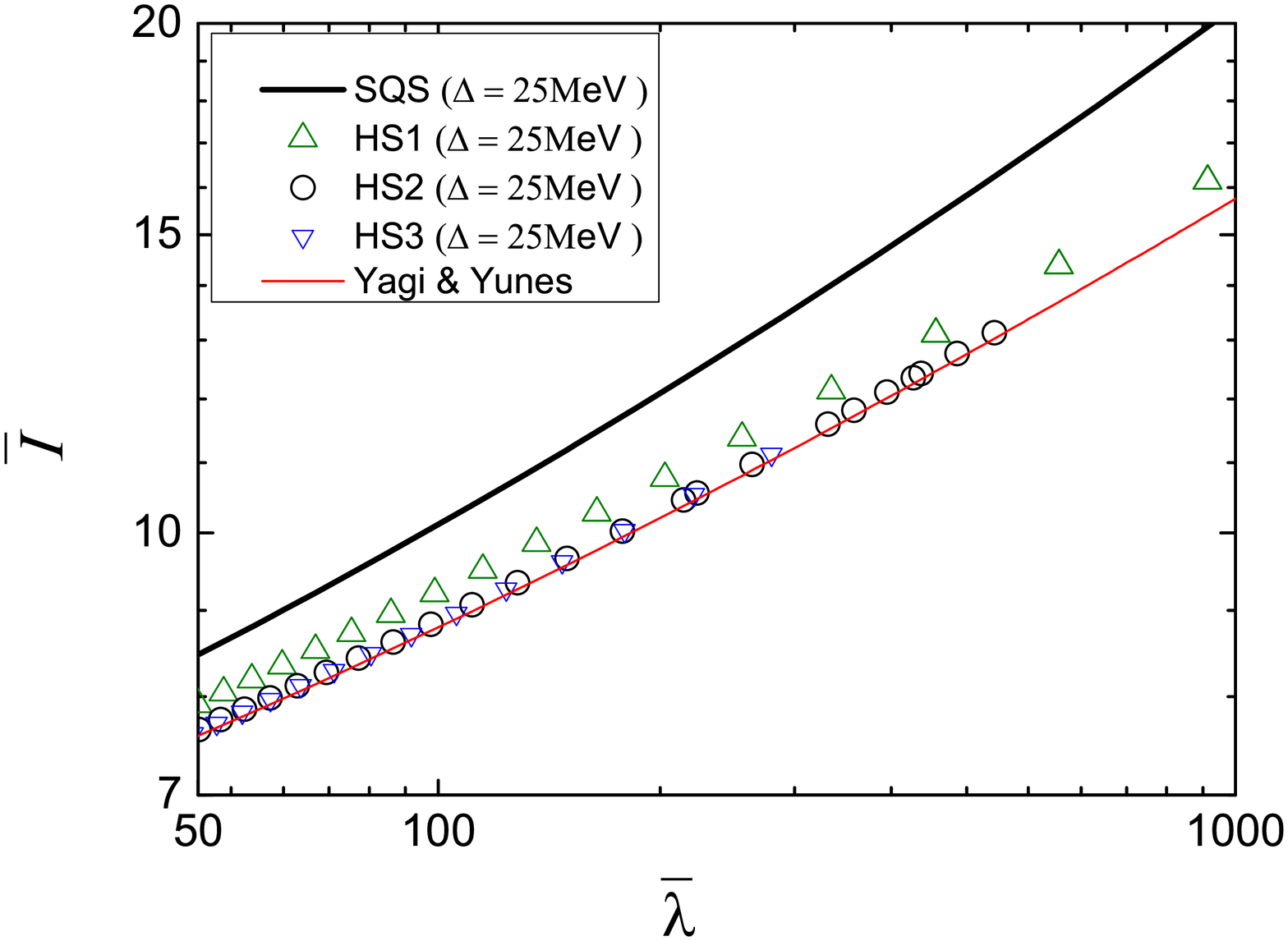}
		\caption{The I-Love relation of the hybrid star models (HS1 - HS3) and solid quark star model (SQS) with gap parameter $\Delta$ = 25~MeV are shown together with the fitting formula of the universal relation in fluid compact stars given by Yagi and Yunes \cite{2013Sci...341..365Y,2013PhRvD..88b3009Y}.}	\label{I_Love_QS_HS}
	\end{figure}
	
	Finally, we end this subsection by showing the I-Love relations of the hybrid stars (HS1 - HS3) together with the solid quark star (SQS) using $\Delta = 25~\text{MeV}$ in Fig.~\ref{I_Love_QS_HS}. It extends our previous study in \cite{2017PhRvD..95j1302L} with more focus on hybrid star models with very different internal structures. The quantity $\bar{I}$ is the normalized moment of inertia, as defined in \cite{2013Sci...341..365Y,2013PhRvD..88b3009Y}, given by 
	\begin{equation}
	\bar{I} = I / M^3,
	\end{equation}
	where $I$ is the moment of inertia and $M$ is the mass of the compact star. The deviation in $\bar{I}$ of HS1 from that of the universal curve for fluid compact stars is around 5\%. Nevertheless, its deviation is still much smaller than that of the solid quark star due to screening effect. 
	Therefore, hybrid star models with a very large solid core like HS1 is in principle distinguishable from pure fluid compact stars using the I-Love relation, if $\bar{I}$ and $\bar{\lambda}$ can be measured to a high accuracy in the future. On the other hand, the screening effect due to the fluid envelope of a hybrid star with a small solid core makes it very difficult to distinguish such a model from a pure fluid model using the I-Love relation alone.
	
	\subsection{Dressed quark stars with a solid CCS quark matter core and a thin layer of nuclear matter crust}
	
\begin{figure}[!t]
		\centering
		\includegraphics[width=8.6cm]{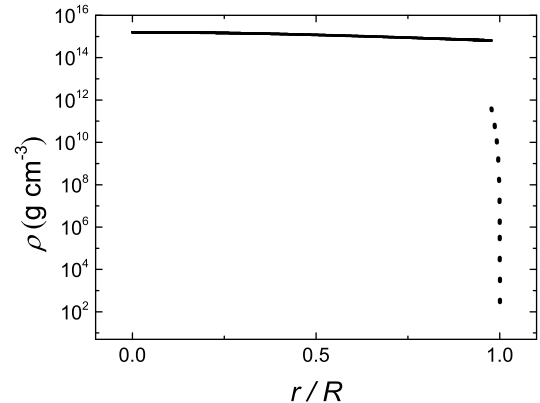}
		\caption{The density profile of the dressed quark star model. The CCS quark matter core (solid lines) occupies more than 95\% of the total radius. The nuclear matter crust is indicated by dotted lines.}	\label{NQS_Stat_Profile}
	\end{figure}

	Another model of quark stars is a two-layer model composed of quark matter dressed in a thin layer of nuclear matter crust \cite{1986ApJ...310..261A}, which is called the `dressed quark star' (e.g., in \cite{Weber:2012ta}) or the `nonbare quark star' \cite{2015ApJ...815...81M}. We would stick to the former term when referring to this model within this paper. Based on the strange matter hypothesis \cite{1984PhRvD..30..272W}, the major component of this model is the absolutely stable strange matter \cite{1986ApJ...310..261A}. At the surface of the quark matter core, a thin layer of nuclear matter with densities below the neutron drip point may exist as a conventional neutron star crust with a layer of electrons of only a few hundreds fermi thick \cite{1986ApJ...310..261A} separating the charged nuclear matter from the quark matter inside. This model is different from the hybrid stars in Subsection~\ref{ssec:BG} since there is no quark matter-nuclear matter transition inside the star. Essentially the quark matter phase is the true ground state and the matter within the star exists in this phase, except that the nuclear matter crust is in a metastable state without direct contact with the quark matter inside. In our investigation of tidal deformation, the thin electron layer is insignificant and we do not include it in the calculation. In contrast to the nuclear matter envelope of hybrid stars, the nuclear matter crust in the dressed quark star has a very low density and there is a density gap of a ratio of $10^3$ to the density of the base of the nuclear matter crust at the interface. Moreover, the nuclear matter crust is a solid with much lower shear modulus than the CCS quark matter inside. The density profile is given in Fig.~\ref{NQS_Stat_Profile}, showing the quark matter core with nearly constant density and the nuclear matter crust with a steep profile.
	
	In the study of torsional pulsation modes of the dressed quark stars \cite{2015ApJ...815...81M}, in which the nuclear matter crust is assumed to be ionic solid, it is found that most of the oscillation mode energy concentrate within the crust since the much more rigid CCS phase quark matter in the interior absorbs only a small faction of energy. The solid nuclear matter crust cracks more easily than the standard neutron star crusts during a glitch. This shows the significance of the nuclear matter crust despite its low mass content compared to the whole star.
	
	We calculate the effect of the thin nuclear matter on the tidal deformability of a dressed quark star. We assume the quark matter to be entirely in the CCS phase. We have employed the same EOS model as SQS in the quark matter core (see Subsection \ref{ssec:BG}). For the layer of nuclear matter, we employ the EOS of \cite{1994A&A...283..313H}, where the matter is assumed to exist as a crust of neutron-rich nuclei. For a nuclear matter crust, we estimate the shear modulus of the crust with the formula \cite{1991ApJ...375..679S} 
	\begin{equation}
	\mu = 0.1194 \frac{n_i (Z e)^2}{R_\text{i}},
	\end{equation}
	where $n_i$ is the number density of ions, $Z$ is the atomic number of the nuclei, $e$ is the electron charge and $R_\text{i}$ is the mean radius of the ions. This gives a shear modulus of $1.47\times10^{28}~\text{erg~cm}^{-3}$ at the base of the crust.
	
	\begin{figure}[!t]
		\centering
		\includegraphics[width=8.6cm]{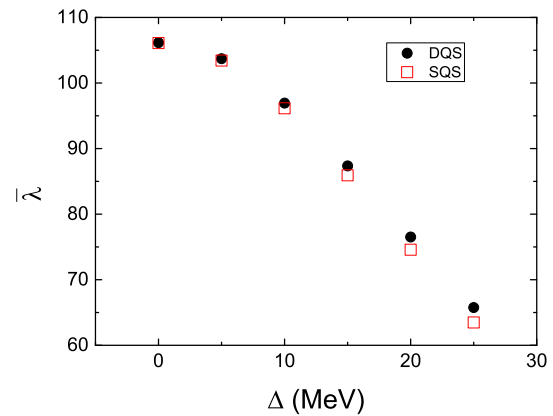}
		\caption{The normalized tidal deformabilities of the dressed quark stars (DQS) and bare solid quark stars  (SQS) with different gap parameters $\Delta$ are plotted. The point with $\Delta = 0~\text{MeV}$ corresponds to the models with fluid quark matter core.}	\label{NQS_Love}
	\end{figure}
	
	The dressed quark star model is labeled as `DQS'. The quark matter-nuclear matter transition is set at a pressure of $7.141\times10^{29}~\text{dyn cm}^{-2}$ so that the bottom of the nuclear crust has the same density as that of the neutron drip point (taken to be $4.2~\times~10^{11}~\text{g~cm}^{-3}$). The quark matter core has a radius of 9.00~km and the nuclear matter crust is 0.21~km thick. We compare the normalized tidal deformabilities of three DQS models with different gap parameters $\Delta$ with those of SQS in  Fig.~\ref{NQS_Love}. We find that the thin layer of solid nuclear matter crust with a low density changes the values of the tidal deformabilities by very little, with less than 5\% for the models with $\Delta = 25~\text{MeV}$. As the gap parameter of the quark matter core increases, the tidal deformability of DQS deviates more from those of SQS models. This is because the effect of elasticity of the solid nuclear matter crust becomes less important when the shear modulus of the CCS phase increases, thus making the nuclear matter crust `fluid-like' in terms of elastic properties. This causes a screening effect similar to the case caused by the fluid envelope in hybrid stars (Subsection~\ref{ssec:BG}).
	
	We also calculate the tidal deformability of dressed quark stars with a fluid nuclear matter crust and compare it with that of dressed quark stars with a solid crust in Table~\ref{NQS_fluid_crust}. The normalized tidal deformability of the model with a fluid crust is labeled as $\bar{\lambda}_\text{fc}$. The tidal deformabilities of the two models agree to within $0.01\%$. 
	
	\begin{table}[!t]
		\centering
		\begin{tabular}{rccc}
			\hline \hline
			\multicolumn{1}{c}{$\Delta$} & 
			\multicolumn{1}{c}{$\bar{\lambda}$}& 
			\multicolumn{1}{c}{$\bar{\lambda}_\text{fc}$}&
			\multicolumn{1}{r}{\% difference (\%)}\rule{0pt}{3ex} \\ \hline
			5 & 103.699 & 103.699  & 0\\
			15 & 87.347 & 87.348  & 1$\times 10^{-3}$\\
			25 & 65.770 & 65.774  & 6$\times 10^{-3}$\\
			\hline
		\end{tabular}%
		\caption{A comparison of the normalized tidal deformabilities of a dressed quark star with a solid nuclear matter crust ($\bar{\lambda}$) and that of a corresponding model with a fluid crust ($\bar{\lambda}_\text{fc}$). The DQS models are fixed at 1.4~$\text{M}_\odot$.}
		\label{NQS_fluid_crust}%
	\end{table}%
	
	In conclusion, the nuclear matter crust of the dressed quark star poses a screening effect on the solid quark matter core. However, it is much less significant than that in a hybrid star. Also, such effect does not depend on whether the nuclear matter crust is solid or fluid as shown in Table~\ref{NQS_fluid_crust}. As we shall see in Sec.~\ref{sec:factors}, the weakness of screening is caused by the large density gap at the core-envelope interface. Hence, the tidal deformability of dressed quark star models with a solid quark matter core in CCS phase is nearly the same as that of bare solid quark star models. In Section~\ref{sec:factors}, we shall demonstrate how the density gap can affect the screening effect.
	
	\subsection{Two-layer quark stars with a fluid CFL quark matter core and a solid CCS quark matter envelope}\label{ssec:2}
	
	If the high density environment of the bare quark star core favors the CFL phase quark matter, the star might be composed of a fluid CFL core and a rigid envelope of CCS quark matter \cite{2014PhRvD..89j3014M}. However, the exact CFL-CCS transition point is unknown and is not possible to be determined without knowing the strange quark mass and the gap parameter $\Delta$. Therefore, we treat the transition point as a free parameter and study a series of quark star models with a fluid CFL core and a solid CCS envelope, focusing on the effect of the envelope on the tidal deformability of the model. This represents an investigation on a different kind of `screening effect' in this model compared to that in the hybrid stars. Specifically, we now ask whether the solid envelope changes the tidal deformability of the two-layer model with a fluid core considerably regardless of its thickness.  
	
	Using the phenomenological EOS of quark matter, Eq.~(\ref{QM_EOS}), we construct three two-layer quark star models, with different transition pressures between the CFL and CCS phases using the parameters $a_4$ = 0.8, $ a_2^{1/2}$ = 100~MeV, $B_\text{eff}^{1/4}$ = 160~MeV as shown in Table~\ref{2layer_QS_models}. We label the models from `CFL-CCS1' to `CFL-CCS3' according to the transition point between the CFL and CCS phases. The mass of each model is fixed at 1.4~$M_\odot$. We have applied the same set of phenomenological EOS parameters to describe the CFL core and the CCS envelope here, assuming that the EOSs of these two phases do not differ significantly. This two-layer model has been previously studied in \cite{2014PhRvD..89j3014M} on the electromagnetic signals emitted from bare quark stars through torsional oscillations. Similar models have also been discussed in \cite{2013PhRvC..88f5801R} for studying the \textit{r}-mode instability of bare quark stars with a transition between a Kaon-condensed CFL phase \cite{2002NuPhA.697..802B,2002PhRvD..65e4042K} and the CCS phase, except that the transition point is determined microscopically by comparing the free energies of these two phases. Among the three two-layer quark stars, CFL-CCS1 has the largest fluid core and CFL-CCS3 has the smallest one.
	
	\begin{table}[!t]
		\centering
		\begin{tabular}{rc|cc}
			\hline \hline
			\multicolumn{1}{c}{EOS} & 
			\multicolumn{1}{c}{$P_t (10^{35}~\text{dyn cm}^{-2})$}& 
			\multicolumn{1}{|c}{${R_c}/{R}$}&
			\multicolumn{1}{c}{$\bar{\lambda}_{\Delta  =25~\text{MeV}}$}\rule{0pt}{3ex} \\ \hline
			CFL-CCS1 & 0.73 & 0.75  & 93.140\\
			CFL-CCS2 & 1.62 & 0.50  & 74.623\\
			CFL-CCS3 & 2.394 & 0.25  & 64.977\\
			\hline
		\end{tabular}%
		\caption{Two-layer quark star models constructed with the EOS parameters $a_4$ = 0.8, $ a_2^{1/2}$ = 100~MeV, $B_\text{eff}^{1/4}$ = 160~MeV are listed. The CFL-CCS transition pressure, $P_t$, is listed for each of the models. The fractional core radii, ${R_c}/{R}$, are also given. For the last two columns, the mass of each model is fixed at 1.4~$M_\odot$ and the normalized tidal deformabilities of the models with the gap parameter $\Delta = 25~\text{MeV}$ are also listed.}
		\label{2layer_QS_models}%
	\end{table}%
	
	\begin{figure}[!t]
		\centering
		\includegraphics[width=8.6cm]{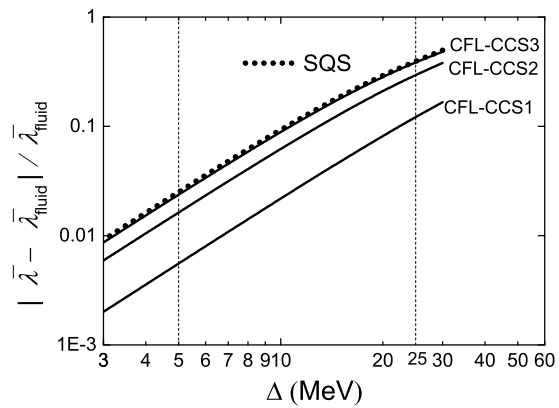}
		\caption{The fractional deviation in normalized tidal deformability against the gap parameter, $\Delta$, of two-layer quark stars CFL-CCS1, CFL-CCS2, CFL-CCS3 and solid quark star SQS. The curve for solid quark star SQS is represented by dots and it nearly overlaps with that of CFL-CCS3. The models are all fixed at 1.4~$M_\odot$. The range of $\Delta$ we consider is bounded by two dashed lines.}	\label{lo:fig3}
	\end{figure}
	
	In Fig.~\ref{lo:fig3}, the fractional deviations in normalized tidal deformability of the two-layer quark star models and that of the bare solid quark star are plotted against the gap parameter. Compared to the hybrid stars in the previous subsection, the deviations for two-layer quark stars are generally larger. The tidal deformability of the two-layer quark star model with $R_c/R = 0.25$, CFL-CCS3, is almost indistinguishable from that of the solid quark star SQS. For comparison, CFL-CCS1, the model with $R_c/R = 0.75$, has a large deviation in tidal deformability from that of SQS for different gap parameter $\Delta$. In general, the tidal deformability of a two-layer quark star with a CCS envelope approaches the value of that of a bare solid quark star as the thickness of the envelope increases, unlike the case of a two-layer model with a fluid envelope and a solid core which shall be discussed in Sec.\ref{sec:factors}. 
	
	\begin{figure}[!t]
		\centering
		\includegraphics[width=8.6cm]{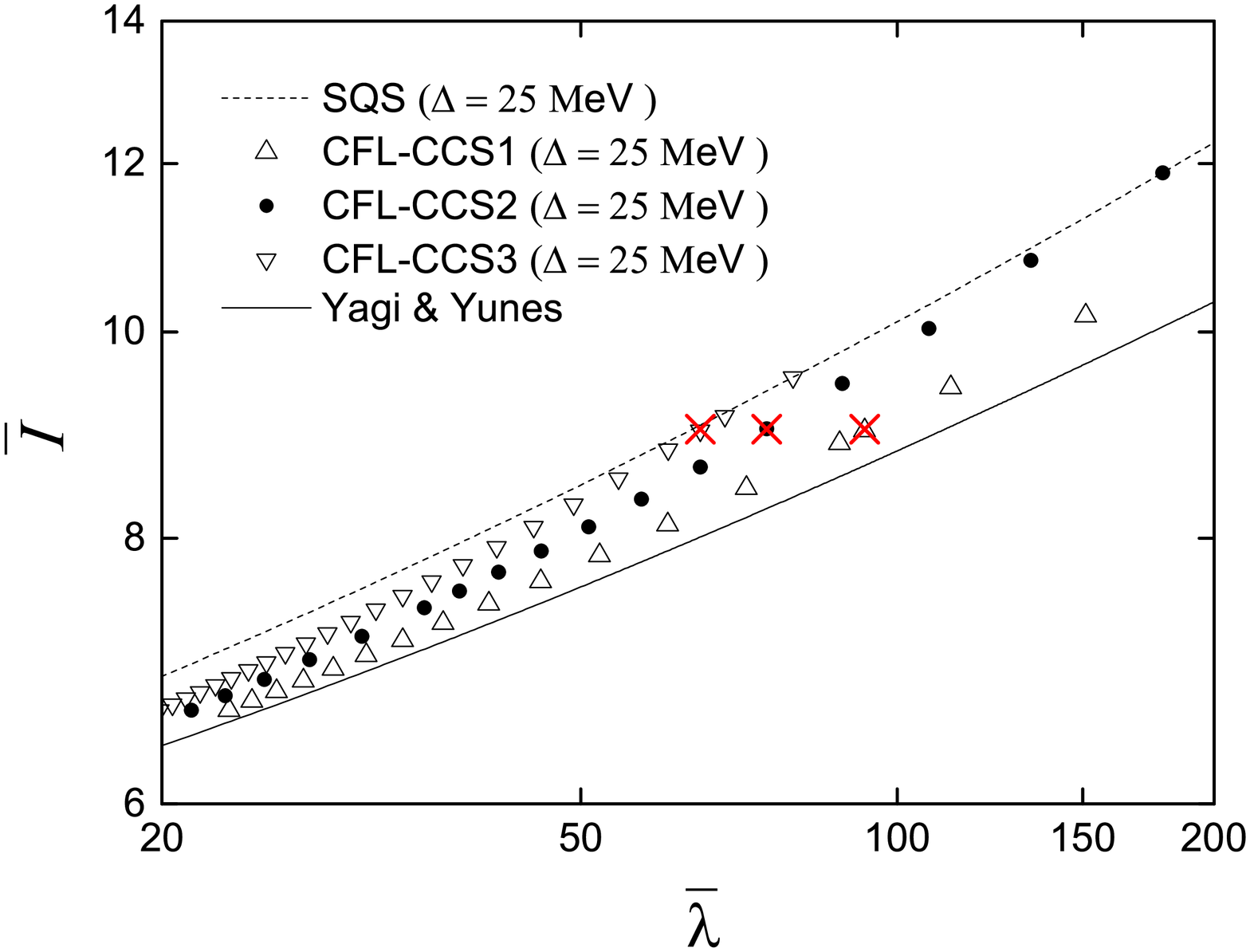}
		\caption{The I-Love relation of the two-layer quark star models tabulated in Table~\ref{2layer_QS_models} and solid quark star models (SQS) with gap parameter $\Delta$ = 25~MeV are shown together with the fitting formula of the universal relation in fluid compact stars given by Yagi and Yunes \cite{2013Sci...341..365Y,2013PhRvD..88b3009Y}. The two-layer quark star models with mass $1.4~M_\odot$ are indicated with red crosses presented in Table~\ref{2layer_QS_models}.}	\label{I_Love_FE}
	\end{figure}
	
	Figure~\ref{I_Love_FE} illustrates the I-Love relations of the two-layer quark star models with different CFL-CCS transition pressures. The gap parameter is fixed at 25~MeV. Since we employ the same EOS for the CFL phase and the CCS phase, these models have the same density profiles as the SQS model. The transition is thus characterized by the position at which the shear modulus changes sharply from zero in the fluid CFL phase to an extremely large value in the solid CCS envelope. The models with mass $1.4~M_\odot$ are marked by red crosses in the figure. As shown in Fig.~\ref{I_Love_FE}, the I-Love relation of the two-layer quark star can deviate from the universal I-Love relation by a significant amount depending on the size of the CFL core. This depends on the transition pressure between the CFL and CCS phases as well as the gap parameter. In contrast to the cases of hybrid stars, the deviation is now more sensitive to the transition point of the fluid-solid interface. This attributes to the difference in behavior of the screening effect caused by a fluid envelope and that caused by a solid envelope. 
	
	\subsection{Summary}\label{ssec:3}
	
	Here we briefly summarize the numerical findings in this section. We have studied the tidal deformabilities of three kinds of two-layer compact star models containing the CCS phase quark matter. We focus on the reduction in the influence of the core on the tidal deformability caused by the outer layer, which we refer to as the screening effect.
	
	\begin{itemize}
		\item Hybrid stars: For our stellar models with a typical density gap at the core-envelope interface which is comparable to the core density, we find that the screening effect is very strong as long as the CCS core size is smaller than about 70\% of the stellar radius. Let us also point out that the rather uniform density profile of the quark matter core in our models also contributes to the strong screening effect in hybrid stars. This shall be discussed in Subsection~\ref{ssec:coresize}.
		\item Dressed quark stars: The solid nuclear matter crust has a screening effect similar to that caused by the fluid nuclear matter envelope of a hybrid star, except that the effect is much weaker. The tidal deformability of this model is slightly higher than that of a bare solid quark star with identical mass as a result of the screening effect of the crust. We shall illustrate in Sec.~\ref{sec:factors} that it is due to the large density gap of the ratio $10^3$ compared to the base of the nuclear matter crust at the core-crust interface.
		\item Two-layer quark stars (CFL-CCS): In contrast to the case of hybrid stars, the screening effect in this model depends more sensitively on the transition point between the solid envelope and fluid core.
		As a result, the I-Love relation for these models can deviate significantly from the universal relation for pure compact stars.
	\end{itemize}
	
	\section{Factors affecting the screening effect} \label{sec:factors}
	The previous results demonstrate the screening effect in two-layer compact star models containing the CCS phase quark matter, together with a study of the general dependence of the effect on the relative thickness between the envelope and inner core. In this section, we shall use a series of `toy models', including polytropic models and incompressible models, to study the factors affecting screening effect in more detail.
	
	For illustration, we use a polytropic model ($P = k \rho^{1 + 1/n}$) and an incompressible model ($\rho = \rho_0$), where $k$ is a constant and $n$ is the polytropic index, $\rho_0$ is a constant. The values of the EOS parameters are fixed at $k = 180$~km$^2$, $n=1$ and $\rho_0 = 10^{15}$~g~cm$^{-3}$ respectively. The shear moduli of the solid phase in both the polytropic model and the incompressible model are set to be a constant value of 2.87$\times 10^{34}$~erg~cm$^{-3}$, which is close to that of the CCS phase quark matter with gap parameter $\Delta = $ 25~MeV. We also employ a two-layer quark star model, labeled as `2-layer QS', constructed with the EOS model given in Subsection~\ref{ssec:BG}, but with a core-envelope transition between CCS quark matter and fluid quark matter at an adjustable radius $R_c$. Again we choose the EOS parameters for quark matter as: $a_4$ = 0.8, $ a_2^{1/2}$ = 100~MeV, $B_\text{eff}^{1/4}$ = 160~MeV.
	
	The screening effect has been studied in Newtonian theory with a two-layer incompressible model featuring a solid core and a fluid envelope of different densities \cite{1979Icar...37..310D,2015Icar..248..109B,2015Icar..258..239B}. It depends on several parameters of the internal structure, including the density gap across the interface, the relative sizes of the envelope and the crust, and the shear modulus of the solid core. We expect the dependence to be similar in compact stars. In the following, we numerically study the effects of these factors within the framework of GR.
	
	\subsection{Core size} \label{ssec:coresize}
	The screening effect in models with a solid core and a fluid envelope is somewhat surprising. Within the linearized theory, we find that the screening effect does not vanish even when the fluid envelope is very thin as long as the density of the envelope is non-zero. It is analogous to the electrostatic shielding effect of a perfect conductor enclosing a dielectric material (see e.g., \cite{Griffiths:1492149}). When a dielectric material is exposed to an external electric field, a polarization is induced within it. However, if it is completely surrounded by a perfectly conducting shell, it does not feel the external electric field since the electric field from the induced charges on the conducting shell surface completely cancels out the external electric field. Such a screening effect is independent of the thickness of the perfect conductor shell. In the case of the tidal deformation problem, a perfect fluid layer responds to the external tidal field with an induced quadrupole moment, which drastically cancels out the external field on the solid core. The case of complete screening can be shown analytically in Newtonian incompressible models (see Appendix~\ref{app2}).
	
	We now study the special features of the screening effect due to the envelope by comparing two different kinds of compact star models: one with a solid core surrounded by a fluid envelope, while the other model is a reverse situation with a solid envelope on top of a fluid core. In these two situations, screening simply means the reduction of the effect of the core on the tidal deformability due to the existence of an outer envelope. To simplify the discussion, we study models without a density gap at the core-envelope interface. We define a screening factor, $\delta \bar{\lambda} / (\delta \bar{\lambda})_{\text{max}}$, of each model, with
	\begin{align}
	\delta \bar{\lambda} &= \Big|\bar{\lambda} - \bar{\lambda}^*\Big|, \\
	(\delta \bar{\lambda})_{\text{max}} &= \bar{\lambda}_{\text{fluid}} - \bar{\lambda}_{\text{solid}},
	\end{align}
	where $\bar{\lambda}$ is the normalized tidal deformability of the two-layer compact star, $\bar{\lambda}_{\text{fluid}}$ and $\bar{\lambda}_{\text{solid}}$ are that of the corresponding single-layer fluid model and solid model respectively. Here, a single-layer model refers to one with the same background profile (e.g., $P(r)$, $\rho(r)$) as the two-layer compact star, except that the whole star is composed entirely of either fluid or solid. $\bar{\lambda}^*$ is that of the reference model, which is a single-layer compact star composed of the same phase as the core of the two-layer model, i.e., either $\bar{\lambda}_{\text{fluid}}$ or $\bar{\lambda}_{\text{solid}}$. Hence, $\delta \bar{\lambda} / (\delta \bar{\lambda})_{\text{max}}$ ranges from 0 (no screening) to 1 and it indicates how strong the screening effect by the envelope on the core is. For instance, when the normalized tidal deformability of the two-layer model with a fluid core and a solid envelope is the same as that of the fluid model, i.e. $\bar{\lambda} = \bar{\lambda}_{\text{fluid}} = \bar{\lambda}^*$, the screening factor vanishes and there is no screening in this situation. 
	
	\begin{figure}[!t]
		\includegraphics[width=8.6cm]{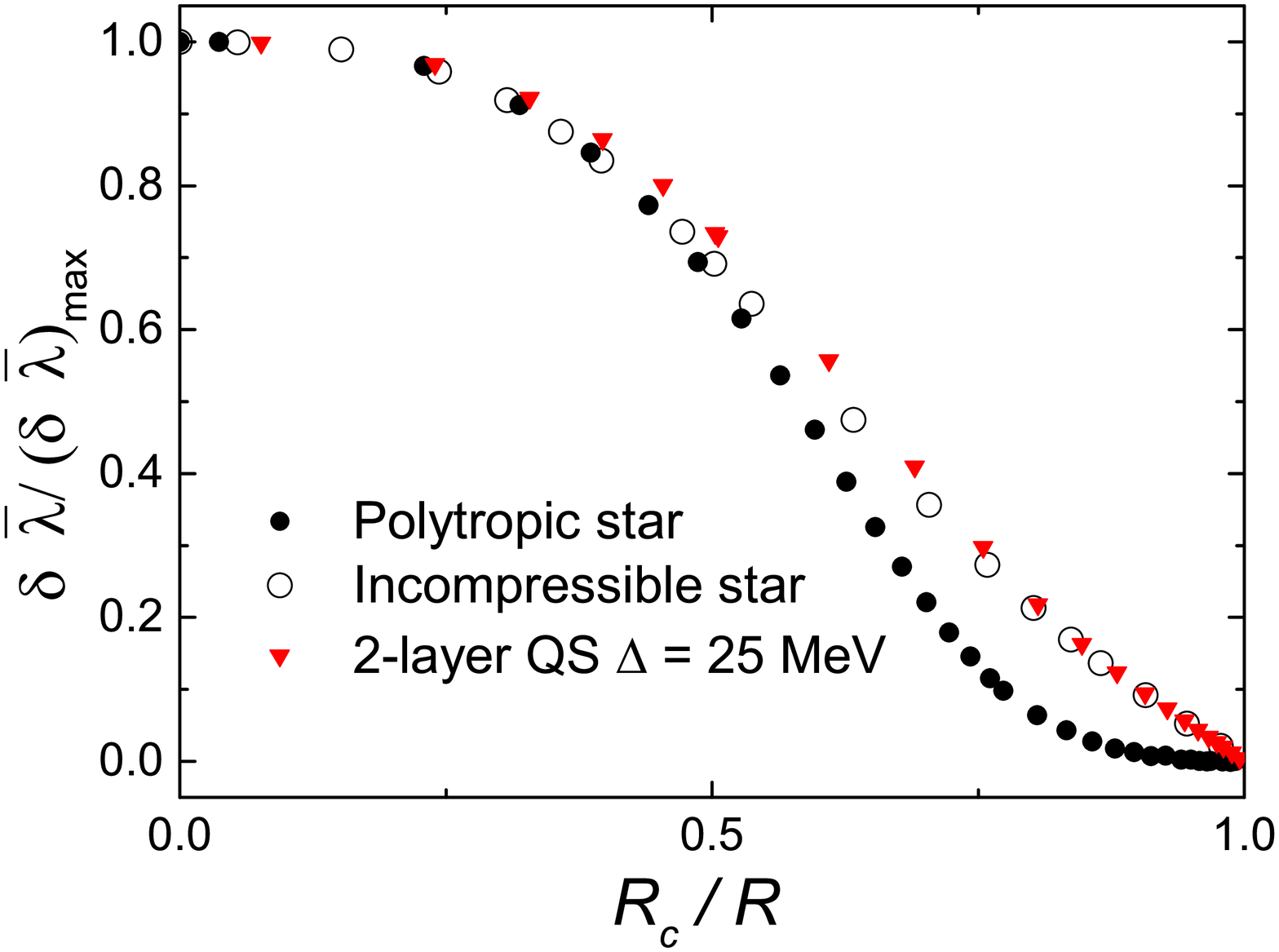}
		\caption{\label{Screening_FE} The screening factor plotted against the fractional core radius of the two-layer models with a fluid core and a solid envelope. The models contain no density gap at the interface.}
	\end{figure}
	In Fig.~\ref{Screening_FE}, the screening factor of different two-layer models with a fluid core and a solid envelope reduces from 1 to 0 as $R_c / R$ increases for all three models. It illustrates a gradual decrease in screening effect in these models as the solid envelope gets thinner. In particular, this situation applies to the case of traditional neutron stars with a thin solid nuclear-matter crust.
	
	\begin{figure}[!t]
		\includegraphics[width=8.6cm]{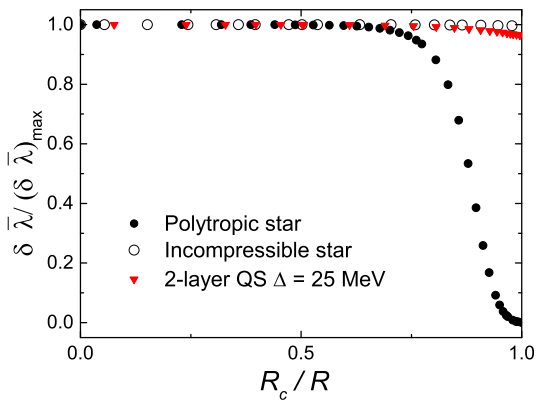}
		\caption{\label{Screening_EF} The screening factor plotted against the fractional core radius of the two-layer models with a solid core and a fluid envelope. The models contain no density gap at the interface.}
	\end{figure}
	
	In Fig.~\ref{Screening_EF}, we observe strong screening in all three models when $R_c / R$ is less than 0.75, with the screening factor extremely close to 1. As $R_c / R$ increases, the screening factor of the polytropic star starts to decrease and eventually reaches zero as $R_c / R = 1$. Meanwhile, the screening factors of incompressible star and quark star stay at a value near 1. The screening effect is still very significant even when $R_c / R$ approaches 1 for the incompressible model and quark star model, both of which have finite surface densities. The two-layer quark star model has a slightly weaker screening effect than the incompressible star. This highlights the special feature of the screening effect of a fluid envelope on a solid core, which is not very sensitive to the thickness of the fluid envelope, especially for stellar models with a finite surface density. Moreover, the screening effect does not vanish as long as the fluid envelope has a significant density compared to the solid core for the incompressible star and quark star. 
	
	From the different behaviors in screening factor between the polytropic model and the other two models when $R_c / R $ tends to 1 , we see that the uniformity of the density profiles also contributes to the screening effect. In the two-layer quark star model, the density at the surface of the solid core is at least a quarter of its central density as $R_c / R $ approaches 1. For comparison, the density of the solid core in a polytropic model can be several orders of magnitude smaller than the central density if the core-envelope interface is close to the surface. This causes a significant drop in the screening effect when $R_c/R$ is larger than about 0.8 as shown in Fig.~\ref{Screening_EF}. This result is similar to the case of hybrid star models studied before where the screening effect is weakened as the solid-core size increases (see Figs.~\ref{Density_HS_14} and \ref{lo:fig2}). 
	
	\subsection{Density gap} \label{ssec:den_gap}
	
	In this subsection, we illustrate the effect of density gap at the core-envelope interface on the screening effect. This can qualitatively explain the large difference between the screening effect in hybrid stars and that in dressed quark stars. While screening effect is observed in dressed quark stars, the effect is not as significant as that in hybrid stars. This is due to the tremendous difference between the ratios of the density gap to the stellar core density at the interface in the two models. In this subsection, we compare the change in the screening factor as we adjust the density gap of a two-layer incompressible model with a solid core and a thin fluid envelope. We calculate the tidal deformabilities of models with an incompressible solid core surrounded by a very thin fluid envelope with adjustable density $\rho_\text{f}$. 
	
	Numerically, it can be implemented by replacing the boundary conditions at the stellar surface of a single layer solid star with the conditions at the core-envelope interface of a two-layer model, and setting $R_c = R$. Although it might not be valid to employ linear theory to calculate the tidal deformations in these models with such a thin fluid layer as the non-linear terms start to dominate when the deformation is comparable to the fluid layer thickness, it still serves as a reference on how the screening effect would depend on the density gap for realistic models with a thicker fluid envelope. Note that it is not possible to isolate the effect of density gap as the only changing factor for models with a thick fluid envelope. Instead, factors like the thickness of the fluid envelope and the non-uniformity of the density profile might dominate over the effect of the density gap. For this reason, we compare the dependence of the screening factor on the density gap in such a thin-envelope limit so that we can isolate the density gap as the only changing parameter without altering the stellar structure. 
	
\begin{figure}[!t]
		\centering
		\includegraphics[width=8.6cm]{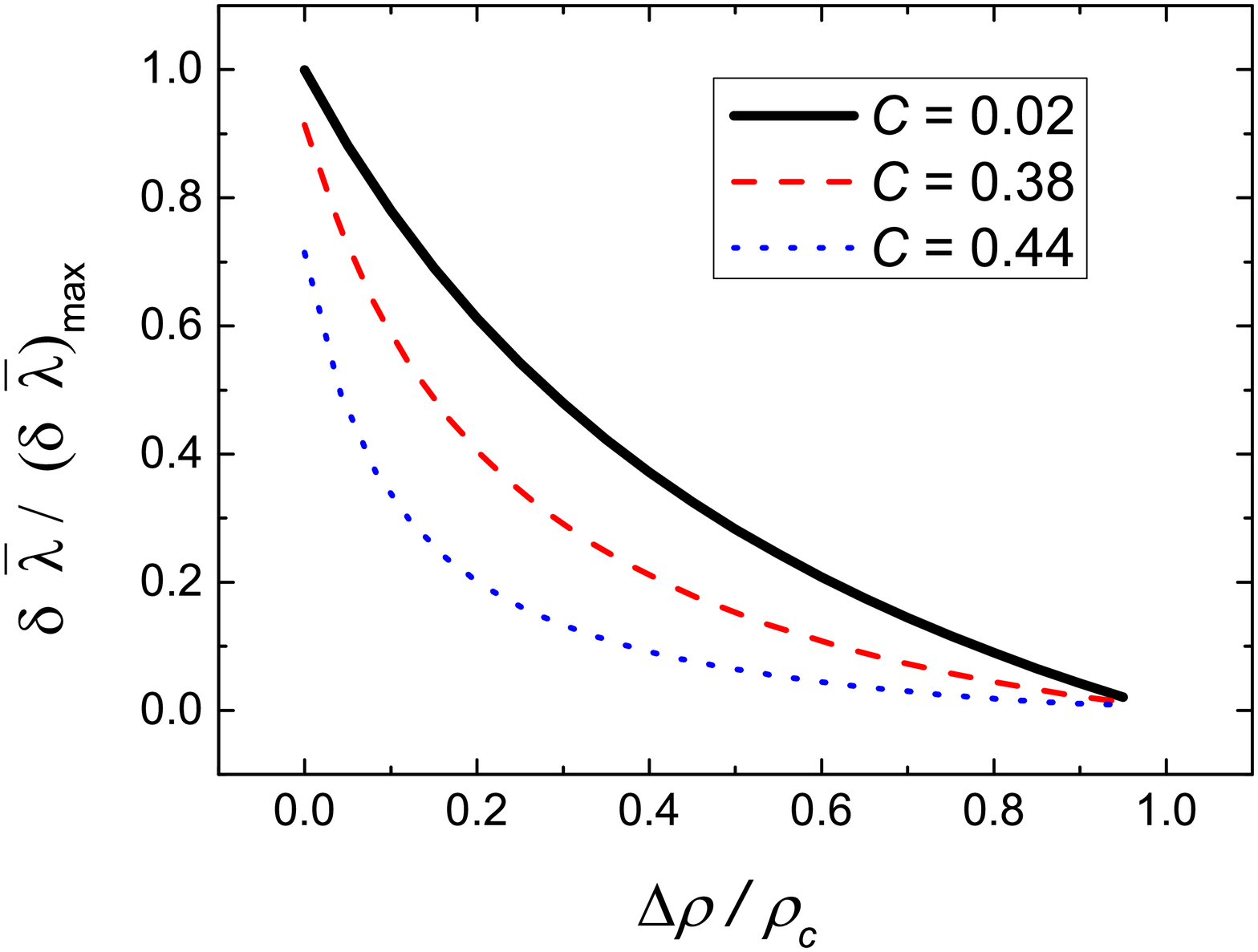}
		\caption{\label{InC2_Screening_drho_v2} The screening factor, ${\delta \bar{\lambda}} / {\big(\delta \bar{\lambda}\big)_\text{max}}$, against ${\Delta \rho} / {\rho_c}$ of two-layer incompressible models with solid core densities of 10$^{15}$~g~cm$^{-3}$ and fluid envelope densities $\rho_\text{f}$. The fluid envelope is set to be very thin so that we can freely adjust the density gap without altering the overall background profile.}
	\end{figure}

	We adjust the density of the fluid envelope $\rho_f$ and plot the screening factor, ${\delta \bar{\lambda}} / {\big(\delta \bar{\lambda}\big)_\text{max}}$, against ${\Delta \rho} / {\rho_\text{c}}$ for three models of different compactness in Fig. \ref{InC2_Screening_drho_v2}, where $\rho_\text{c}$ is the uniform solid-core density and ${\Delta \rho} / {\rho_\text{c}}$ is defined by
	
	\begin{equation}
	\frac{\Delta \rho}{\rho_\text{c}} = \frac{\rho_\text{c} - \rho_\text{f}}{\rho_\text{c}}.
	\end{equation}
	Although the fluid envelope of our model is very thin, they still pose a rather strong screening effect if the density gap is small. We can see the significant decrease in screening effect when $\Delta \rho$ is increased. In Fig.~\ref{InC2_Screening_drho_v2}, the screening factor ${\delta \bar{\lambda}} / {\big(\delta \bar{\lambda}\big)_\text{max}}$ decreases gradually towards 0 as the density gap increases for models with different compactness. This indicates that as long as $\rho_\text{f}$ is not too low, the thin fluid envelope can still have a significant screening effect on the solid core. Fig.~\ref{InC2_Screening_drho_v2} shows clearly how the screening effect is weakened as the density gap is increased. 
	
	\subsection{Compactness}
	
	Fig.~\ref{InC2_Screening_drho_v2} also shows that, for a given density gap, the effect of relativity also weakens the screening effect. To further study the effect of relativity, we use the incompressible model without a density gap at the core-envelope interface. It is known in the Newtonian limit that the fluid envelope of a two-layer incompressible model without a density gap can perfectly screen out the external tidal field on the solid core (See Appendix~\ref{app2}). This has been studied previously in \cite{1979Icar...37..310D,2015Icar..248..109B,2015Icar..258..239B}. In the following, we numerically show that increasing the compactness can reduce the effect of screening in our general relativistic incompressible model.
	
\begin{figure}[!t]
		\centering
		\includegraphics[width=8.6cm]{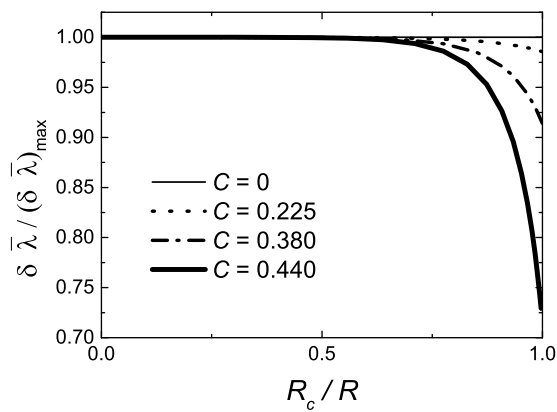}
		\caption{\label{InC1_screening_v2} The screening factor, ${\delta \bar{\lambda}} / {\big(\delta \bar{\lambda}\big)_\text{max}}$, against the fractional core radius of incompressible stars with different compactness.}
	\end{figure}

	In Fig.~\ref{InC1_screening_v2}, the screening factor of two-layer incompressible models with a solid core and a fluid envelope is plotted against the fractional core radius. The models have different compactness as indicated in the legend of Fig.~\ref{InC1_screening_v2}. It is note that the effect of relativity becomes important at the high end of $R_c/R$. In particular, for $R_c/R \gtrsim 0.7$, the screening factor decreases rapidly as the compactness increases. While the incompressible model in Newtonian limit ($C=0$) has a screening factor equal to 1 within numerical accuracies, the one with maximum compactness $(C=0.44)$ has a screening factor of around 0.75 as $R_c / R \rightarrow 1$. This demonstrates that the effect of GR causes a reduction in screening effect.
	
	Figure~\ref{InC1_screening_v2} gives a qualitative understanding on how strongly GR can influence the screening effect. From the figure, the reduction in screening effect is very small, with a magnitude within 2\%, for typical compact stars of compactness around 0.2. Hence, the effect of GR on the screening effect is not significant for typical compact star models.
	
	\section{Implications from GW170817}
	\label{sec:GW170817}
	
	On 17 August 2017, the Advanced LIGO and Virgo network made the first direct detection of the gravitational wave from a binary compact star merger, GW170817  \cite{PhysRevLett.119.161101}. The correlated electromagnetic signals in different frequency bands were also detected (see \cite{2041-8205-848-2-L12} and the references therein). From the LIGO and Virgo observations, an upper bound on the normalized tidal deformability of a $1.4~M_\odot$ star is approximated to be 800 in a low-spin scenario \cite{PhysRevLett.119.161101}. This upper bound has already been used to put constraints on EOSs (see, e.g., \cite{2018ApJ...852L..29R,PhysRevLett.120.172703,PhysRevD.97.083015,PhysRevD.97.084038,2018ApJ...857L..23R,2018ApJ...852L..25R,2017ApJ...850L..34B,PhysRevLett.120.172702,2018RAA....18...24L,Abbott:2018exr}).
	
\begin{figure}[!t]
		\centering
		\includegraphics[width=8.6cm]{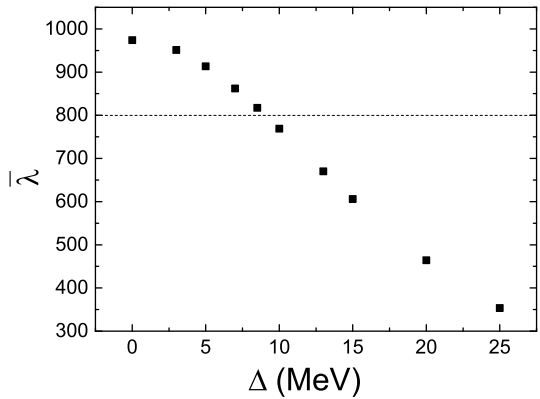}
		\caption{\label{GW170817M14} The normalized tidal deformability, $\bar{\lambda}$, of a $1.4~M_\odot$ solid quark star with $a_4 = 0.7$, $a_2^{1/2} = 0$ and $B_\text{eff}^{1/4} = 135~\text{MeV}$. As the gap parameter $\Delta = 0~\text{MeV}$, i.e., for a fluid quark star, $\bar{\lambda}$ exceeds the upper bound of 800 \cite{PhysRevLett.119.161101}. For $\Delta > 9~\text{MeV}$, $\bar{\lambda}$ is within the upper bound. }
	\end{figure}

	We have shown that elasticity can reduce the tidal deformability significantly in certain models containing CCS phase, including the solid quark stars \cite{2017PhRvD..95j1302L}, dressed quark stars with a CCS core and quark stars containing a CFL core and a CCS envelope.
	Hence, the constraints on EOS parameters for fluid quark stars discussed in \cite{PhysRevD.97.083015}
	need to be reconsidered if the quark matter is in a crystalline phase such as the CCS phase 
	that we focus in this paper or the quark-cluster model proposed in \cite{2003ApJ...596L..59X}. 
	As an illustration, we take our quark-matter EOS (see Eq.~(\ref{QM_EOS})) with parameters
	$a_4 = 0.7$, $a_2^{1/2} = 0$ and $B_\text{eff}^{1/4} = 135~\text{MeV}$. The normalized tidal deformability of a $1.4 M_\odot$ fluid quark star constructed with this EOS is $\bar\lambda = 974$, which is ruled out by the upper bound $\bar\lambda = 800$ obtained from GW170817. However, for a solid quark star with the same mass and EOS, the normalized tidal deformability can decrease below 800 if the gap parameter is larger than 9~MeV as shown in Fig.~\ref{GW170817M14}.

	During inspiral, the solid layer(s) of the compact stars can be melted if the stress from the tidal field is too large. Postnikov \textit{et al.} \cite{2010PhRvD..82b4016P} estimated the frequency of the gravitational wave signals at which the solid quark matter crust reaches the threshold strain and breaks. For a $1.4~M_\odot$ quark star with a solid crust with shear modulus $4\times 10^{32}~\text{erg~cm}^{-3}$, the frequency at the breaking point is about $12~\text{Hz}$, which is outside the best sensitivity region of Advanced LIGO and Virgo at around 100 Hz \cite{RILES20131,PhysRevD.93.112004}. However, the shear modulus of the CCS phase quark matter strongly depends on the gap parameter which spans a wide range from 5 to 25~MeV phenomenologically. Assuming the gap parameter is 25~MeV, the shear modulus would be about $2\times 10^{34}~\text{erg~cm}^{-3}$ for a $1.4~M_\odot$ solid quark star with EOS parameters $a_4 = 0.7$, $a_2^{1/2} = 0$,  $B_\text{eff}^{1/4} = 135~\text{MeV}$. The corresponding breaking frequency would then be about $180~\text{Hz}$ following the estimation method in \cite{2010PhRvD..82b4016P}. As a result, the quark matter may still be in the solid phase when the emitted gravitational wave signal is detected during the inspiral if the gap parameter is near its theoretical upper bound. 
	
	On the other hand, we expect that the tidal deformability of a hybrid star model containing a CCS quark matter core would be very close to that of a hybrid star with a fluid core due to screening effect, assuming that the solid core size is smaller than about 70\% of the stellar radius (see Subsection~\ref{ssec:BG}). 
	As a result, if a hybrid star EOS model is ruled out by the observational upper bound on the tidal deformability, the conclusion will hold no matter whether the quark matter is in a fluid or solid state, assuming that a large solid core comparable to the stellar radius is not favored in nature.
	
	\section{Conclusion}
	\label{sec:conclude}
	
	In this paper, we study the tidal deformability of compact star models containing the extremely rigid CCS phase quark matter. We have presented a formulation to determine the tidal deformability of two-layer compact stars with a solid component. Comparing to previous work on this subject (e.g., \cite{2011PhRvD..84j3006P}), our formulation is written in terms of a different set of matter variables so that the resulting equations can be compared directly to their Newtonian counterparts. We have applied our formulation to study four different compact star models: (1) solid quark stars composed entirely of CCS quark matter \cite{2017PhRvD..95j1302L}; (2) hybrid stars with a nuclear matter fluid envelope on top of a CCS quark-matter core \cite{2008PhRvD..77b3004I,2009PhRvD..79h3007K,2013PhRvD..88l4002L}; (3) dressed solid quark stars with a thin nuclear matter solid crust \cite{2015ApJ...815...81M}; and (4) two-layer quark stars with a fluid CFL quark-matter core surrounded by a CCS quark-matter envelope \cite{2014PhRvD..89j3014M}. 
	We focus on the screening effect on the tidal deformability due to the envelope of various 
	two-layer compact star models, which screens off the influence by the elastic or fluid property of the core. 
	
	Our results show that the screening effect in hybrid stars is strong as long as the size of the solid quark-matter core is less than about 70\% of the stellar radius. For instance, the fractional deviation in the normalized tidal deformability of a hybrid star, with a solid core with radius about half of the stellar radius from
	the corresponding pure fluid model is below 1\%.
	
	On the other hand, the screening effect in dressed solid quark stars with a thin nuclear-matter crust featuring a large density gap at the core-crust interface is very weak. Further analysis in Subsection~\ref{ssec:den_gap} shows that the large density gap is the reason for the weakness of the screening effect. In other words, if the density gap is zero, the screening effect would become so strong that the tidal deformability of the model would deviate a lot from that of a solid quark star with the same background profile.
	
	We have also found that the screening effect in two-layer quark star models with a 
	fluid CFL core surrounded by a CCS solid envelope is different from hybrid stars in terms of the dependence on the thickness of the core and envelope. Compared
	to the case of hybrid stars, the screening effect of two-layer quark stars is more sensitive to the position of transition between the fluid core and solid envelope. 
	
	Besides, we also investigate how the screening effect in two-layer compact stars is affected by the core size, the density gap at the core-envelope interface and the compactness of the stars. First, we adjust the core size of two-layer models without a density gap at the interface to study its influence on the screening effect. For models with a fluid core and a solid envelope, the screening effect gradually changes with the core size. The screening factor, defined in Sec.~\ref{sec:factors}, reduces from 1 to 0 (no screening) as the core size increases from 0 to the stellar radius. On the contrary, the screening effect of models with a solid core and a fluid envelope show a much weaker dependence on the core size. We also find that for models with a rather uniform density profile, the screening factor remains close to 1 for any core size between 0 and the stellar radius, which indicates strong screening regardless of the core size. For polytropic models with a solid core and a fluid envelope, the screening factor remains close to one for core size less than 0.75 of the stellar radius.
	
	We also show that the screening factor of a two-layer incompressible model with a solid core and a thin fluid envelope reduces gradually to 0 as the density gap at the core-envelope interface increases from 0 to 100\% of the core density. This indicates that the density gap at the interface is an important factor to affect the screening effect. It specifically explains the weak screening effect in dressed quark stars.
	
	The effect of GR on the screening effect is also studied by varying the compactness of our stellar models. A slight reduction on the screening factor is found on two-layer incompressible models as the compactness increases from the Newtonian limit (i.e., compactness equals 0), to the highly relativistic case (compactness equals 0.44). However, the reduction in screening factor is not significant as long as we are considering the typical range of compactness of around 0.2.
	
	Our numerical investigation suggests that the screening effect depends crucially on the detailed stellar structure such as the core size, composition (fluid or solid state) of the core and envelope, and the density gap at the core-envelope interface. 
	
	Finally, we have demonstrated how quark star models which
	are ruled out by the observation limits on the tidal deformability 
	obtained from GW170817 \cite{PhysRevD.97.083015} can be revived if the entire quark star is in a CCS phase instead of a fluid phase. 
	This illustrates how the crystalline phase of quark matter might come into play when one tries to use the information on the tidal deformability obtained from gravitational wave observations to put constraints on quark-matter EOS models. Our study also advocates that the tidal deformability not only provides us information on the EOS, but may also give insights into the multi-layer structure and elastic properties of compact star models composed of CCS quark matter. With the expectation that more compact star mergers will be observed in the coming decades, the possibility of using the observed gravitational wave signals to constrain the various models of deconfined quark matter will become very promising.
	
	\appendix
	\section{Regular solutions near origin}\label{app1}
	The set of perturbation equations in solid has a regular singular point at the origin. To determine the regular solutions near the origin, we expand the perturbation variables in power series of $r$ about the origin following the approach by Finn \cite{1990MNRAS.245...82F}:
	\begin{equation} \label{app1_expansion}
	Q_n(r) = r^{\alpha_n}\left( Q_n^{(0)} + Q_n^{(2)} r^2 + O(r^4) \right),
	\end{equation}
	where $n$ ranges from 1~to~6 and $Q_1(r)$~to~$Q_6(r)$ represents the set of perturbation variables $\left\{ W(r), Z_r(r), V(r), Z_\perp(r), H_0(r), J(r) \right\}$ in corresponding order. Substituting the above expressions into the perturbation equations (Eqs.(\ref{fn_eqt1})-(\ref{fn_eqt6})), the leading power dependences of the regular solutions, $\alpha_1$ to $\alpha_6$, are given by $\left\{ l, l-2, l, l-2, l, l-1 \right\}$.
	
	The background variables such as $\rho$ and $P$ are also expanded in $r$ and are expressed as 
	\begin{align}
	\rho(r) =& \rho_0 + \rho_2 r^2 + O(r^4),\\
	P(r) =& P_0 + P_2 r^2 + O(r^4).
	\end{align}
	The twelve coefficients of the perturbed variables, $Q_n^{(0)}$ and $Q_n^{(2)}$ with $n = 1$, ..., $6$, are related by nine independent constraints. This permits three independent regular solutions at the origin. We derive the explicit forms of the constraints and give them as follows:
	\begin{widetext}
		\begin{align}
		W^{(0)} =& l V^{(0)}, \label{static_origin_regular_01} \\ 
		\bigg[\gamma P_0 (l+3) + \frac{\alpha_1}{3}(l+9)\bigg]W^{(2)} =& - \frac{1}{2}\Big[\rho_0 + (3\gamma + 1) P_0 \Big]H_0^{(0)} + \frac{4}{9}\Bigg\{18 \pi \alpha_1 P_0 \bigg(1-6\gamma \bigg)  + 9 l \pi P_0\big(\rho_0 + P_0 - 2\alpha_1 \big)  \label{static_origin_regular_02} \\
		&+ \pi \rho_0\Bigg[3l\Big(\rho_0 + (1 - 2\gamma)P_0 \Big) + 2 \bigg(l(3l-4)-9\bigg)\alpha_1\Bigg]\Bigg\}V^{(0)} \nonumber\\
		& + \Big[\gamma P_0 l + \frac{\alpha_1}{3}(l-6)\Big](l+1) V^{(2)}, \nonumber\\
		(4l + 6) H_0^{(2)} =& - 4 \pi \bigg\{-\frac{l(2l +3)-15}{3}\rho_0 + 3(l+3-9 \gamma)P_0 \bigg\} H_0^{(0)}  \label{static_origin_regular_03} \\
		& -\frac{32 \pi ^2}{3} \Bigg\{3 P_0l\bigg[\Big(3 + \frac{1}{{c_s}^2}\Big)(\rho_0 +P_0) - 4 \alpha_1\bigg]  \nonumber\\
		&+\rho_0 l \bigg[(1+\frac{1}{{c_s}^2})(\rho_0 +P_0) -6 \gamma P_0 - \alpha_1 \bigg]+ 12\alpha_1\Big[- \rho_0 + (3 - 9 \gamma)P_0\Big] \Bigg\}V^{(0)}  \nonumber\\
		&  -8\pi \Big[\rho_0 + (1+3\gamma)P_0\Big] \Big[-(l+3)W^{(2)} + l(l+1)V^{(2)}\Big], \nonumber\\
		Z_r^{(0)} =& -2 \alpha_1 l (l - 1) V^{(0)},  \label{static_origin_regular_04} \\
		Z_r^{(2)} =& -\Big[\frac{8\pi}{3} l \rho_0 \alpha_3 + 16 \pi \big(\alpha_3 + 2\alpha_2\big) \alpha_1 \Big]V^{(0)} - \frac{1}{2} \big(\alpha_3 + 2\alpha_2\big) H_0^{(0)} \label{static_origin_regular_05}\\   
		& - \Big( l \alpha_3+ \alpha_3 + 2 \alpha_2 \Big) W^{(2)} + L_1  \alpha_2 V^{(2)}, \nonumber\\
		Z_\perp^{(0)} =& -2 \alpha_1 (l - 1) V^{(0)},  \label{static_origin_regular_06} \\
		Z_\perp^{(2)} =& -\alpha_1 \Big[ -\frac{8 \pi}{3} \rho_0 \big(l -2\big) V^{(0)}  + W^{(2)} + l V^{(2)}\Big],  \label{static_origin_regular_07}\\
		J^{(0)} =& l H_0^{(0)} - 8 \pi \big(\rho_0 + P_0\big) V^{(0)},\label{static_origin_regular_08}\\
		J^{(2)} =& \big(l+2\big)H_0^{(2)} - \frac{8 \pi}{3}\bigg[ 16\pi \alpha_1 \big(\rho_0+3P_0\big)+ 3 l (\rho_2+P_2)\bigg]V^{(0)} + 8 \pi \big(\rho_0+P_0\big)W^{(2)}. \label{static_origin_regular_09}
		\end{align}
	\end{widetext}
		The variables $\gamma$ and $c_s^2$ are respectively defined by
	\begin{align}\label{static_origin_regular_10}
	\gamma &= \frac{\rho_0 + P_0}{P_0} \Big(\frac{d P}{d \rho}\Big)_0,\\
	{c_s}^2 &= \Big(\frac{d P}{d \rho}\Big)_0 \; ,
	\end{align}
	where $\Big(d P / d \rho\Big)_0$ is the leading term in the Taylor series expansion of $ dP / d\rho$ in $r$ about the origin. The expansion coefficients of $\rho$ and $P$ are related by
	\begin{align}
	P_2 =& -\frac{4 \pi}{3} (\rho_0 + 3P_0) (\rho_0 + P_0),\\
	\rho_2 =&  \frac{P_2}{{c_s}^2} .
	\end{align}
	
	The three independent regular solutions are constructed by choosing three independent sets of $\big(H^{(0)}$, $V^{(0)}$, $V^{(2)}\big)$ in the above nine constraints, e.g., $\left(1, 0, 0\right)$, $\left(0, 1, 0\right)$ and $\left(0, 0, 1\right)$. Next, we find the values of $W^{(0)}$, $W^{(2)}$ and $H_0^{(2)}$ from Eqs.~(\ref{static_origin_regular_01})-(\ref{static_origin_regular_03}). This allows us to directly determine the remaining coefficients from the above relations.
	
	The above set of constraints on the regular solutions are consistent with those of the zero frequency limit in the Newtonian pulsation problem (e.g., \cite{1975GeoJ...41..153C}) at the Newtonian limit. Note that although Finn \cite{1990MNRAS.245...82F} has derived the constraints for the regular solutions around the center of the solid core in the relativistic case, he mistakenly imposed an additional constraint on the shear stress and in turn causes $W^{(0)} = V^{(0)} = 0$, which makes the solutions inconsistent with those in the Newtonian limit.
	
	In a fluid core, there is only one regular solution of the form~\cite{2008ApJ...677.1216H}:
	\begin{equation}
	H_0(r) = a_0 r^l - a_0 \bigg[\frac{2\pi}{2l + 3} \Big(5 \rho_0 + 9 P_0 + \frac{\rho_0 + P_0}{{c_s}^2}\Big)\bigg] r^{l +2},
	\end{equation}
	where $a_0$ is an arbitrary constant.
	
	\section{Perfect screening in Newtonian incompressible models}\label{app2}
	The screening effect of the tidal deformability of a Newtonian incompressible star with a solid core and fluid envelope has been studied analytically in \cite{1979Icar...37..310D,2015Icar..248..109B,2015Icar..258..239B}. In this appendix, we shall illustrate the simplest case of screening where the effect of the external tidal field on the core is completely screened off by the fluid envelope of a Newtonian model with uniform density. 
	
	In the Newtonian case, the motion of a mass element is governed by the Poisson's equation, conservation of momentum and continuity equation
	\begin{align}
	\nabla^2 \Phi &= 4 \pi \rho, \\
	\frac{\partial \vec{v}}{\partial t} + \vec{v} \cdot \nabla \vec{v} &= \frac{1}{\rho} \nabla \cdot \boldsymbol{\sigma} - \nabla \Phi, \\
	\nabla \cdot (\rho \vec{v}) &= \frac{\partial \rho}{\partial t},
	\end{align}
	where $\Phi$ is the gravitational potential, $\vec{v}$ is the velocity of the mass element, $\boldsymbol{\sigma}$ is the stress tensor in isotropic solids given by \cite{landau1986theory}
	\begin{equation}
	\boldsymbol{\sigma} = \gamma P \nabla \cdot \vec{u} \; \boldsymbol{I} + \mu \bigg[\nabla \vec{u} + (\nabla \vec{u})^{\text{T}} \bigg] - \frac{2 \mu}{3} \nabla \cdot \vec{u} \; \boldsymbol{I},
	\end{equation}
	with $\boldsymbol{I}$ being the identity matrix, $\vec{u}$ being the displacement vector and $\gamma$ being the adiabatic index. The perturbed scalar quantities, $\delta Q$, are expanded in the basis of spherical harmonics:
	\begin{equation}
	\delta Q(r, \theta, \phi) = \sum_{lm} \delta Q_{lm}(r) Y_{lm}(\theta, \phi),
	\end{equation}
	where $\delta Q_{lm}(r)$ is the radial component of the term in the multipole expansion of order $\{l,m\}$. In the following, we shall drop the subscripts $\{l,m\}$ on these functions as we focus on a particular order.
	
	The gravitation potential perturbation is governed by the perturbed Poisson equation (e.g., \cite{cox1980theory})
	\begin{equation} \label{Newt_Poissona}
	\nabla^2 \Big[\delta \Phi(r) Y_{lm}(\theta, \phi) \Big] = 4 \pi \delta \rho(r) Y_{lm}(\theta, \phi) = 0.
	\end{equation} 
	Note that in our case of an incompressible model, $\delta \rho = 0$. Hence, the gravitational potential depends only on the overall shape of the star (i.e., the interface/surface boundary condition) and the external tidal field. 
	
	We have assumed the elastic stress to contribute only in the perturbation level. The momentum conservation equation is derived using the standard strain tensor for isotropic solid and Hookean strain-stress relation. The spheroidal component (of even parity) of the deformation vector of a mass element is denoted by $\vec{\xi} = \xi_r(r) Y_{lm}(\theta, \phi) \hat{r} + \xi_\perp(r) r \nabla Y_{lm}(\theta, \phi)$. The linearized momentum conservation equation for spheroidal deformations is hence given by
	\begin{equation} \label{Newt_Solid_Momentuma}
	\nabla \left[\left(\frac{\delta P}{\rho} + \delta \Phi\right)Y_{lm}(\theta, \phi)\right] + \frac{\mu}{\rho}\nabla \times \nabla \times \vec{\xi} = 0. 
	\end{equation}
	The perturbed continuity equation for spheroidal modes is given by (e.g., \cite{cox1980theory})
	\begin{equation} \label{Newt_Continuitya}
	\nabla \cdot \left(\rho \vec{\xi}\right) = \delta \rho(r) Y_{lm}(\theta, \phi) = 0.
	\end{equation}
	
	Eq.~(\ref{Newt_Poissona}) does not depend on the other perturbed quantities such as $\vec{\xi}$ and can be solved alone to give a regular solution
	\begin{equation} \label{key}
	\delta \phi(r) = A r^l,
	\end{equation}
	where $A$ is some arbitrary constant. Hence, Eqs.~(\ref{Newt_Solid_Momentuma}) and (\ref{Newt_Continuitya}) are left with two regular solutions, i.e., two undetermined constants. As a result, the perfect screening of the external tidal field on the solid core is easily realized by considering the fact that the two boundary conditions at the core-envelope interface are homogeneous in $\xi_r$ and $\xi_\perp$ when the density gap across the interface is zero:
	\begin{equation} \label{radial_stress_cont}
	\Big[\Delta P + 2\mu \frac{d \xi_r}{dr}\Big]_{R_{c-}} = \Big[\Delta P\Big]_{R_{c+}},
	\end{equation}
	\begin{equation} \label{tangential_stress_cont}
	\mu \Big[ r \frac{d \xi_\perp/r}{dr} + \frac{\xi_r}{r} \Big]_{R_{c-}} = 0,
	\end{equation}
	where we use the subscripts `$+$' and `$-$' to denote different sides of an interface at radius $R_c$, e.g., $R_{c+} = \lim\limits_{\delta \rightarrow 0^+} \big(R_c + \delta\big)$.
	
	It can be shown that Eq.~(\ref{radial_stress_cont}) gives a homogeneous equation of $\xi_r(R_{c-})$ and $\xi_\perp(R_{c-})$ when there is no density gap, which in turn gives the trivial solution $\xi_r = \xi_\perp \equiv 0$ within the solid core when combined with the other homogeneous relation, Eq.~(\ref{tangential_stress_cont}), since $\xi_r(R_{c-})$ and $\xi_\perp(R_{c-})$ are left with two degrees of freedom. This leads to the result $\bar{\lambda} \equiv 1 / (2 C^5)$, independent of the solid core radius, which exactly equals the normalized tidal deformability of a Newtonian fluid incompressible star. This is the so-called perfect screening, where the effect of the external tidal field on the solid core is completely screened off by the deformed fluid envelope.
	
	
	
	\addcontentsline{toc}{chapter}{Reference}
	\bibliographystyle{apsrev4-1-etal}
	\bibliography{database}
	
\end{document}